\documentclass[12pt,preprint]{aastex}

\usepackage{emulateapj5}
\newcommand{\et}{et al.}

\newcommand{\fv}{F_{\rm var}}
\newcommand{\fvar}{F_{\rm var}}

\newcommand{\xte}{{\it RXTE}}

\newcommand{\xmm}{{\it XMM-Newton}}

\newcommand{\Msun}{\hbox{$\rm\thinspace M_{\odot}$}}
\slugcomment{}
\shorttitle{Markowitz et al.}
\shortauthors{Mkn 766 X-ray Variability}
\begin{document}
\title{The Energy-dependent X-ray Timing Characteristics of the Narrow Line Seyfert 1 Mkn 766}

\author{A.~Markowitz\altaffilmark{1,2}, I.~Papadakis\altaffilmark{3}, P.~Ar\'{e}valo\altaffilmark{4}, T.J.~Turner\altaffilmark{1,5}, L.~Miller\altaffilmark{6}, J.N.~Reeves\altaffilmark{1,7} 
\altaffiltext{1}{X-ray Astrophysics Laboratory, Code 662, NASA/Goddard Space Flight Center, Greenbelt, MD 20771, USA; agm@milkyway.gsfc.nasa.gov}
\altaffiltext{2}{NASA Post-doc Research Associate}
\altaffiltext{3}{Physics Department, University of Crete, P.O.\ Box 2208, 71003 Heraklion, Greece}
\altaffiltext{4}{School of Physics and Astronomy, University of Southampton, Southampton S017 1BJ, UK}
\altaffiltext{5}{Department of Physics, University of Maryland Baltimore County, 1000 Hilltop Circle, Baltimore, MD 21250, USA}
\altaffiltext{6}{Department of Physics, University of Oxford, Denys Wilkinson Building, Keble Road, Oxford OX1 3RH, UK}
\altaffiltext{7}{Department of Physics and Astronomy, Johns Hopkins University, Baltimore, MD 21218, USA} }

\begin{abstract}   %%%% now at 195

We present the energy-dependent power spectral density (PSD) 
and cross-spectral properties of Mkn 766, obtained from 
combining data obtained during an {\it XMM-Newton} observation 
spanning six revolutions in 2005 with data obtained from an 
{\it XMM-Newton} long-look in 2001. The PSD shapes and rms-flux 
relations are found to be consistent between the 2001 and 2005 
observations, suggesting the 2005 observation is simply a 
low-flux extension of the 2001 observation and permitting us 
to combine the two data sets. The resulting PSD has 
the highest temporal frequency resolution for any AGN PSD 
measured to date. Applying a broken power-law model yields
break frequencies which increase in temporal frequency with 
photon energy. Obtaining a good fit when assuming energy-independent 
break frequencies requires the presence of a Lorentzian at 
$4.6\pm0.4 \times 10^{-4}$ Hz whose strength increases with 
photon energy, a behavior seen in black hole X-ray binaries.
The cross-spectral properties are measured; temporal 
frequency-dependent soft-to-hard time lags are detected in this 
object for the first time. Cross-spectral results are
consistent with those for other accreting black hole systems. The 
results are discussed in the 
context of several variability models, including those based on 
inwardly-propagating viscosity variations in the accretion disk.

\end{abstract}

\keywords{galaxies: active --- galaxies: Seyfert --- X-rays: galaxies --- galaxies: individual (Mkn 766) }

\section{Introduction}

Seyfert active galactic nuclei (AGNs) and stellar-mass black hole 
X-ray binary systems (XRBs) both exhibit rapid, aperiodic X-ray 
variability that likely originates in the innermost regions of 
these compact accreting objects. The dominant X-ray radiation is 
generally thought to be inverse Comptonization of soft seed photons 
by a hot corona (e.g., Shapiro \et\ 1976, Sunyaev \& Titarchuk 1980), 
though the exact geometry is uncertain and numerous configurations 
have been invoked (e.g., Zdziarski \et\ 2003). The X-ray variability 
can be characterized by fluctuation power spectral density functions 
(PSDs) which reveal the "red-noise" nature of the variability at 
relatively high temporal frequencies in both AGNs and XRBs. Modeling 
of the broadband PSDs of XRBs usually utilizes some combination 
of one, two, or more Lorentzians, which tend to dominate in the 
so-called low/hard energy spectral state, plus a broadband noise 
component characterized as $P(f) \propto f^{-1}$, which tends to 
dominates in the high/soft state (e.g., in Cyg X-1; see Axelsson \et\ 
2005). Quasi-periodic oscillations, or QPOs, are detected in some 
XRBs, usually those in the so-called Very High spectral state (e.g.,
McClintock \& Remillard 2006).

Broadband PSDs have also been constructed for AGNs, allowing  
modeling of the overall PSD shape. Specifically, 'breaks' in 
the power-law PSD models on time scales of a few days or less 
have been identified for roughly a dozen AGN (e.g., Uttley, 
M$^{\rm c}$Hardy \& Papadakis 2002; Markowitz \et\ 2003; Vaughan, 
Fabian \& Nandra 2003b, hereafter VFN03; M$^{\rm c}$Hardy \et\ 
2004, 2005), with power-law slopes $\lesssim -2$ above the break 
and $\sim -1$ below it. No evidence for QPOs 
has been found in AGN PSDs yet. One reason for this may that
the insufficient sampling of the observed AGN light curves 
so far does not allow one to achieve a high enough temporal frequency resolution 
(Vaughan \& Uttley 2005). Nonetheless, the overall similarity in 
Seyferts' and XRBs' broadband PSD shapes and scaling of PSD break 
time scale with black hole mass (e.g., Markowitz \et\ 2003, 
M$^{\rm c}$Hardy \et\ 2004), as well as the presence of a linear 
relation between flux and absolute rms variability (e.g., 
Uttley \& M$^{\rm c}$Hardy 2001, Edelson \et\ 2002), support 
the notion of similar X-ray variability mechanisms 
being present in both classes of accreting compact objects. That is, 
Seyferts and XRBs may simply be scaled versions of each other in 
black hole mass and X-ray variability time scale.

Further support for this picture comes from observations of the 
PSD slope above the break to flatten with increasing energy.
Previous studies of both XRBs (e.g., Nowak \et\ 1999a, Lin 
\et\ 2000) and Seyferts (Nandra \& Papadakis 2001, VFN03, 
M$^{\rm c}$Hardy \et\ 2004, and Markowitz 2005; see also 
Leighly 2004) have suggested that the PSD slope flattens 
as photon energy increases. There has been no evidence yet 
for break frequency to change significantly with photon 
energy in either AGN or XRBs during a single observation.

The observed cross-spectral properties, namely the coherence 
function (Vaughan \& Nowak 1997) and time lag spectrum, are also 
similar between Seyferts and XRBs. The coherence is generally 
seen to be close to unity over a wide range of temporal 
frequencies in both AGNs and XRBs, usually dropping significantly
below unity at frequencies above the PSD break (e.g., 
Nowak \et\ 1999a, 1999b in XRBs; VFN03, M$^{\rm c}$Hardy \et\ 2004 
and Markowitz 2005 in AGN). Relatively harder X-rays are seen to 
generally lag those at softer X-ray energies, with the lag 
between bands increasing with increasing energy separation. 
However, lags are are also observed to increase with decreasing 
temporal frequency (e.g., Miyamoto \& Kitamoto 1989, Miyamoto \et\ 
1991, Nowak \et\ 1999a in XRBs), with time lags usually 
$\lesssim 1-10\%$ of the Fourier time scale. In the low-hard state 
of Cyg X-1, the time lags increase in a step-like fashion 
(e.g., Nowak 2000), which can be explained if the lag at each step 
is associated with individual Lorentzian components in the 
broadband PSD. Obtaining time lags via the cross-spectrum requires 
extremely high quality data with sufficient sampling and number of 
points. Results for AGN so far have yielded time lag spectra with 
many fewer points than in XRBs, but also showing the general 
increase in time lags as temporal frequency decreases 
(Papadakis, Nandra \& Kazanas 2001, VFN03, M$^{\rm c}$Hardy
\et\ 2004, Markowitz 2005, and Ar\'{e}valo \et\ 2006a).

Mkn 766 is a well-studied Narrow-Line Seyfert 1 (NLSy1) , and thus
belongs to a class of objects in which large-amplitude rapid 
X-ray variability is commonly observed. Timing properties 
of Mkn 766 have been studied previously by Vaughan \& Fabian 
(2003; hereafter VF03), who measured the broadband PSD, finding 
a break from steep PSD power-law slope of $\lesssim -2.5$ 
to $\sim -1$ near $ 5 \times 10^{-4}$ Hz. VF03 also observed
the slope above the break to flatten from $\sim -2.7$ at soft X-rays to 
$\sim -2.2$ at hard X-rays; however, they saw no obvious 
trends in break frequency with energy. VF03 were the first 
to measure the coherence in Mkn 766. In addition, Vaughan 
\et\ (2003a) observed the linear rms-flux relation in Mkn 766.

In this paper, we present the timing properties of Mkn 766
using the enormous wealth of data obtained from uninterrupted 
\xmm\ observations in 2005 and 2001. We find that the variability properties
of the source during both observations were identical. We therefore combined
the 2001 and 2005 light curves to construct power- and cross-spectra with the 
highest frequency resolution 
in log space seen an in AGN PSD so far. 
The rest of this paper is organized as follows:
$\S$2 describes the data reduction. $\S$3 demonstrates 
similarities in the variability properties of the 2001 and 2005 
observations. $\S$4 describes measurement of, and model fits to, 
the energy-dependent PSD made from combining all the 2001 and 
2005 data. $\S$5 describes the cross-spectral properties. The 
results are discussed in $\S$6 in the context of phenomenological 
variability models. Finally, a brief summary is given in $\S$7.

In one of two companion timing papers, Papadakis \et\ (in prep.; 
hereafter Paper II) combine the 2001 and 2005 \xmm\ light curves with
long-term light curves obtained from \xte\ monitoring. That paper concentrates
on study of the low-frequency PSD shape of Mkn 766 to probe 
similarities to the broadband PSD shapes of XRBs. In the other paper, 
Ar\'{e}valo \et\ (in prep.; Paper III) continue the study of the
energy-dependent variability properties via Fourier-resolved
spectroscopy.

\section{Observations and Data Reduction}

Mkn 766 was observed by \xmm\ on 2005 May 23 -- 2005 June 3,
over revolutions 999--1004. This paper uses data taken with the
European Photon Imaging Camera (EPIC), which consists of one pn CCD
back-illuminated array sensitive to 0.15--15 keV photons
(Str\"{u}der \et\ 2001), and two MOS CCD front-illuminated arrays
sensitive to 0.15--12 keV photons (MOS1 and MOS2, Turner \et\ 2001).
Data from the pn were taken in Small Window mode; data from each
MOS were taken in Large Window mode. The medium filter was used for all detectors. 

Light curves were extracted using SAS version 6.5 using standard
extraction procedures. Data were selected using event patterns 0--4 for the pn
and 0--12 for each MOS. Hot, flickering or bad pixels
were excluded. Source light curves were extracted from
circular regions of radius 40$\arcsec$; background light curves were
extracted from circles of identical size, centered $\sim 3\arcmin$
away from the source. We checked for pile-up; there was no significant pile-up
in the pn and at most a few percent pile-up in each MOS.

The 2--12 keV pn light curve was analyzed for background flares; 
any data where the pn background rate exceeded 0.2 ct s$^{-1}$ were
excluded. Most data excluded occurred at the beginnings or ends of 
revolutions (e.g., the final 18 ksec of revolution 999 data,
the final 17 ksec of revolution 1003, and the final 13 ksec of 
revolution 1004). This screening also introduced a 10 ksec gap into 
the revolution 1003 light curve. Because such large gaps tend to reduce 
the statistical significance of variability parameters derived over 
the full duration, the light curve was split into two smaller light 
curves, hereafter designated 1003A and 1003B, for a total of seven 
virtually uninterrupted light curves. All other gaps introduced by 
background screening were small ($<$ a few ksec) and in the analyses 
below, fluxes during gaps were linearly interpolated from adjacent 
points. 

Data were extracted over the 0.2--12 keV (total), 0.2--0.5 
keV (soft), 0.5--1.1 keV (medium), and 1.1--12 keV (hard) bands
(also referred to as T, S, M and H bands, respectively, below). 
The S, M and H band limits were chosen such that all three bands had 
approximately the same count rates. Consequently, the variability due 
to Poisson noise (see $\S$4 and 5) was roughly the same level, 
mitigating any effect that the difference in count rates or 
Poisson noise between bands could have on analysis. Light curves 
from the three EPIC cameras were summed, using only data taken when 
all three cameras were in operation simultaneously; this yielded a 
total exposure time (summed over all light curves) of 442 ksec. 
Given the wide bands used here, the differences between pn and MOS response 
shapes have a negligible effect; the average pn and MOS photon 
energies for each band differed only by small amounts. In addition, 
Mkn 766 is a soft-spectrum source, and the inclusion of the MOS data 
compared to the pn data alone greatly improved signal/noise in the
M and H bands. Light curves were binned to 60 s;
they are shown in Figure 1. The summed T, S, M and H band light curves
have mean count rates (after background subtraction) of 
13.44, 4.39, 4.39, and 4.93 ct s$^{-1}$, respectively.
We use these 
summed light curves in all analyses below except the rms-flux relation
and $\fvar$ spectra, where only pn data are used.

Mkn 766 was also observed during revolution 265
from 2001 May 20 -- 2001 May 21.
The pn and MOS2 were both in Small Window mode; 
the MOS1 was in Timing mode and is not considered further.
Data were extracted from the pn and MOS2
using the same procedure and bands as for the 2005 data.
No significant pile-up was found in the pn or MOS2.
This reduction yielded a 107.1 ksec light curve
summed over the pn and MOS2
The summed T, S, M and H band light curves
have mean count rates (after background subtraction) of 
26.85, 9.18, 9.34, and 8.12 ct s$^{-1}$, respectively.
The pn light curve is presented in VF03.

\section{Basic Variability Properties}

To quantify variability amplitude as a function of energy,
we constructed $\fv$ spectra for the 2005 observation 
using the pn data and bin sizes of 6 ksec, and using the formulation 
of Vaughan \et\ (2003a). 
Figure 2 shows the $\fv$ spectrum over the entire observation, and 
during the lowest- and highest-flux revolutions; the plots confirm
relatively high levels of variability, as expected for a NLSy1.
The strongest variability amplitudes are found near 1--2 keV in both the 
2001 (VF03) and 2005 $\fv$ spectra, a result similar to e.g., 
VFN03 for MCG--6-30-15.
The results corroborate the principal component analysis of 
Miller \et\ (2006), who demonstrate that the 
variability of the continuum component
in the soft and hard bands is diluted by the presence of less-variable
components, namely the Compton reflection spectrum in the hard band
and by the soft excess and narrow emission lines in the soft band.

Compact accreting black holes display ``weakly non-stationary''
behavior in that the mean and variance both show scatter over
time, although both the underlying PSD shape and the
expectation value of the fractional variability amplitude $<$$\fv$$>$ 
are expected to remain constant in time, e.g, over time scales shorter than decades to centuries
in AGN.  A constant $<$$\fv$$>$ in flux is equivalent 
to the linear rms-flux relation;
Vaughan \et\ (2003a) previously established the presence of both
a linear rms-flux relation and constant $<$$\fv$$>$ with flux
in Mkn 766 using the 2001 data. 

We use the enormous wealth of data from the 2005 observation to revisit this issue. 
The average 0.2--12 keV flux was $3 \times 10^{-11}$ erg cm$^{-2}$ s$^{-1}$ (e.g., Turner \et\ 2006), 
about half that of the 2001 observation, $ 6 \times 10^{-11}$ erg cm$^{-2}$ s$^{-1}$ (e.g., Pounds \et\ 2003),
allowing us to test if the rms-flux relation has persisted over a four year time scale
and can extend to lower fluxes. We measured variability amplitudes in the T-band light curve
over short time scales, using only pn light curves to facilitate comparison between 2001 and 2005.
We calculated the flux, $\fv$, and the rms amplitude $\sigma_{\rm rms}$
every 2000 seconds (subtracting off the variance due to Poisson noise).
Light curves with $\leq$10 points were discarded.
Flux, rms amplitude, and $\fv$ points for all seven light curves
from 2005 were combined; points were sorted according to
flux, and binned in groups of 10. 
Results are shown in Figure 3.
Both observations are consistent with a linear 
$<$$\sigma_{\rm rms}$$>$-flux relation, as shown in top panel,
and with $<$$\fv$$>$ being constant in flux, as shown in the bottom panel.
Moreover, the 2001 and 2005 relations are roughly consistent with each other, indicating
that characteristics of the underlying variability process have not significantly
changed since 2001, and the 2005 observation is
merely a low-flux extension of the 2001 observation.
The positive x-intercepts of the best-fit 
$<$$\sigma_{\rm rms}$$>$-flux relations
suggest the presence of a constant component whose magnitude 
has not changed significantly between the two observations;
this qualitatively corroborates the principal component analysis of Miller \et\ (2006).

We performed another test for strong non-stationarity using the PSD 
shapes; as described in the Appendix, the PSD shapes from 2001 and 2005 
are consistent with each other.
Having established that the variability processes are consistent between the
2001 and 2005 observations, 
we combine the 2001 and 2005 light curves in all analyses below 
to maximize the temporal frequency resolution.
The difference in mean count rates (e.g., summed over pn + MOS2 in 2001
compared to pn + MOS1 + MOS2 in 2005) is irrelevant 
since light curves are mean-subtracted and mean-normalized 
in all analyses below.

Virtually identical results are obtained when using the 2005 data only,
though below we present results from
the combined 2001 and 2005 binned PSD, which
features $\sim 25\%$ higher temporal
frequency resolution in log space.

\section{The Energy Dependence of the High-Frequency PSD}

The high-quality (uninterrupted and highly variable) light curves 
allows us to create a very high quality PSD for Mkn 766; 
in particular, the temporal frequency resolution 
allows very tight constraints on break frequency.
$\S$4.1 describes how the PSDs were constructed;
$\S$4.2 describes model fits to the PSD shapes.

\subsection{PSD Construction}

The construction of the PSDs is summarized here; for 
further details on PSD construction, see
e.g., Uttley \et\ 2002, Markowitz \et\ 2003, VFN03, and references 
therein. Unlike Paper II, we do not  
include \xte\ Proportional Counter Array (PCA) monitoring data, since
the PCA bandpass does not go below 2 keV and
this paper's aim is to explore the PSD behavior across as wide an energy range as possible.
First, for each of the seven light curves from 2005 and for the 2001 light curve, 
the mean was subtracted. Periodograms were then calculated separately for each of the eight
light curves using a Discrete Fourier Transform
(DFT; e.g., Oppenheim \& Shafer 1975). 
The PSD normalization of van der Klis (1997) was used.
All periodogram points from the eight light curves were
combined and sorted in temporal frequency.
Following Papadakis and Lawrence (1993),
the periodogram was binned in log space in groups of 25.
This process yielded a PSD for each of the T, S, M and H bands,
each with a usable temporal frequency range of $ 2.57 \times 10^{-5} - 8.27 \times 10^{-3} $ Hz.
Power at temporal frequencies above $\sim 3 \times 10^{-3}$ Hz were
dominated by the white noise power due to Poisson noise;
this observed level of power agreed well with predicted values
calculated as $P_{\rm Psn} = 2 (\mu + B) \mu^{-2}$, where $\mu$
and $B$ are the exposure-weighted net source and background rates, respectively. 
The T-band PSD $P(f)$ is plotted in Figure 4a (in $f - P(f)$ space).
The S, M, and H-band PSDs are plotted in Figures 5a and 5b 
(both in $f - f\times$$P(f)$ space). 

\subsection{PSD Model Fits}

We now discuss model fits to the broadband PSD shape.
Monte Carlo simulations (e.g., Uttley \et\ 2002) are frequently employed
in PSD measurement when it is necessary to account for PSD measurement
distortion effects and to assign errors to poorly-sampled PSD points.
However, for the present PSDs, we expect no aliasing since each of the individual
light curves were continuously sampled and large gaps were excluded.
The effects of red noise leak from temporal frequencies lower than those
sampled here are likely small given the measured PSD shape, 
and at any rate are assumed not to vary significantly with photon energy.
%%%%(at most 0.1 in slope). 
Finally, because each PSD bin contains a sufficient number of 
unbinned periodogram points (25), PSD errors are
normal, removing the need for Monte Carlo simulations to determine
PSD errors. All model fitting was done in log space.
We used least-squares fitting (e.g., Bevington 1969)
to determine the PSD amplitude. 
Errors below correspond to $\Delta\chi^2$ = 2.71.

\subsubsection{Unbroken and Broken Power-law Model Fits}

We first fit an unbroken power-law of the form 
$ P(f) = A (f/f_0)^{-\beta} + P_{\rm Psn} $ 
where the normalization $A$ is the PSD amplitude at the (arbitrary) frequency $f_0$. 
$\beta$ is the power law slope; e.g., positive values of $\beta$ denote
red-noise PSD slopes. $P_{{\rm Psn}}$ is the constant 
level of power due to Poisson noise, kept fixed in all fits 
at the predicted values, which are calculated to be
0.16, 0.44, 0.48 and 0.44 Hz$^{-1}$ for the T, S, M and H bands, respectively.
$P_{{\rm Psn}}$ was added to the model as opposed to being subtracted 
from the data, to avoid the possibility of obtaining unphysical 
negative powers. Values of $\beta$ were tested in increments of 0.02.
Results for the best-fit model are listed in Table 1;
residuals to the fits are shown in Figures 4b (T-band) and 5c (S, M, and H bands).
As per VF03, the fits are quite poor and
signal the need for a break in the PSD model.

We next tried fitting broken power-law models.
In AGN PSDs, it has not always been clear if the breaks are
sharp or instead follow a slow bend.
We fit a sharply-broken power-law model of the form
\[P(f)= \left\{ \begin{array}{ll}
           A(f/f_{\rm b})^{-\gamma} + P_{\rm Psn},  & f \le f_b \\
           A(f/f_{\rm b})^{-\beta} + P_{\rm Psn},   & f > f_b \end{array}  
\right. \]
Here, $\gamma$ and  $\beta$ are the power-law slopes
below and above the break frequency $f_{\rm b}$; $A$
is the PSD amplitude at $f_{\rm b}$. We also tested
a slowly-bending power-law model, 
\[P(f)= A f^{-\gamma} ( 1 + (f/f_{\rm b})^{\beta-\gamma})  + P_{\rm Psn} \]
Here, $\gamma$ and $\beta$ are the power-law slopes far below or above,
respectively, the rollover frequency $f_{\rm b}$. $A$ is 
the PSD normalization.  Values of log($f_{\rm b}$)
were tested in increments of 0.01; values of $\beta$ and $\gamma$ were tested in increments of 0.02.
Results for the best-fit models with $\beta$, $\gamma$, $f_{\rm b}$ and $A$ free
are listed in Table 2; and the residuals are shown in Figures 3c and 3d (T-band) and Figures
5d and 5e (S, M, and H bands). 

These fits are significantly better (all $> 4.2\sigma$ confidence in an F-test)
compared to the unbroken power-law model fits.
Both sharply-broken and slowly-bending model shapes yield similar fits.
There is no obvious trend in either $\beta$ or $\gamma$ with photon energy in either case.
We note that $\gamma$ is similar to the PSD power-law slope $\lesssim 10^{-3.5} $ Hz
obtained from fitting the \xmm\ and long-term \xte\ data, as shown in 
Paper II. For either shape, though, $f_{\rm b}$ for the S band is significantly 
lower by $\sim 0.2-0.5$ in the log (depending on the model used) 
compared to $f_{\rm b}$ for the M or H bands.

We explored if it was significant to keep $\beta$ and $\gamma$ 
thawed in the fits. We repeated the fits, with
$\beta$ and $\gamma$ frozen at the average value of the S, M and H-band 
fits. Results are listed in Table 2 and the best-fit models are
plotted in Figures 5a and 5b.
The sum of S, M and H $\chi^2$/$dof$ (degrees of freedom) 
increased from 574.9/534 to 584.9/540 for the sharply-broken model, 
and from 577.6/534 to 588.3/540 for the slowly-bending model.
It was thus not significant at greater than the 87$\%$ confidence level
(in an F-test) to thaw $\beta$ and $\gamma$.
With $\beta$ and $\gamma$ frozen, $f_{\rm b}$, which has typical errors
of 0.05 in the log, is seen to increase with increasing photon energy
(e.g., errors do not overlap at the 90$\%$ confidence level
between the M and H bands).

Refitting the PSDs with $\beta$, $\gamma$ and $f_{\rm b}$ 
all kept frozen at their average values
between the S, M and H bands
yields a total (S+M+H) $\chi^2$/$dof$ of 604.2/543 
(sharply-broken) and 615.9/543 (slowly-bending).
Compared to the model fits with just $\beta$ and $\gamma$ frozen,
we find that it is significant at 
$>$99.99$\%$ confidence in an F-test (for either model) 
to keep $f_{\rm b}$ thawed in the fits.

% SIN, all parms free     0.42   0.01
% SIN, b & g frozen       0.16   0.11 
% SLO, all parms free     0.55   0.02
% SLO, b & g frozen       0.17   0.17
%    AVG for all 4:       0.33   0.08

We conclude that the data are not good enough to allow us distinguish, 
formally, between a sharp break or a slow rollover in the broken power-law 
models. However, the data are of high enough quality to conclusively 
demonstrate a rather unexpected result, one not observed 
previously in the PSDs of either AGN or XRBs: 
the break frequency increases with photon energy. 
This increase in $f_{\rm b}$ with photon 
energy holds regardless of whether 
we use sharply-broken or slowly-bending power law models and whether the 
power-law slopes in these models are frozen or thawed in the fits. 
Furthermore, the data are consistent with a broadband PSD shape that is 
identical for all bands in that the slopes above and below the break 
are energy-independent; this also has not been previously demonstrated.

Depending on whether 
the sharply-broken or slowly-bending model is used,
and assuming $\beta$ and $\gamma$ are energy-independent, we find that  
from the S to M bands, $f_{\rm b}$ increases by 0.08$\pm$0.08 or 0.05$\pm$0.09 in the log;
from the M to H bands, $f_{\rm b}$ increases by 0.13$\pm$0.07 or 0.16$\pm$0.06 in the log; and
from the S to H bands, $f_{\rm b}$ increases by 0.21$\pm$0.08 or 0.21$\pm$0.08 in the log.
In other words, the change in 
$f_{\rm b}$ from the S to H bands is a 2.6$\sigma$ effect. However, this is a conservative estimate, since the errors in 
Table 2 are 90$\%$, not 1$\sigma$, so the significance of this result is probably larger than 2.6$\sigma$.
The average photon energy for each band,
weighted for instrument count rate and exposure time,
is 0.36, 0.75, and 3.9 keV for the S, M and H bands, respectively.
The increase in $f_{\rm b}$ between the S and H bands is a factor of $\sim$1/5 times the ratio
of the average photon energies.

\subsubsection{The Case for a Lorentzian peak in the PSD of Mkn 766}

Previous PSD papers have reported that $\beta$ flattens as photon energy increases
while $f_{\rm b}$ is energy-independent. We fit the S, M and H bands PSDs with the same sharply-
or slowly-bending power-law models, but with $\gamma$ and $f_{\rm b}$ assumed to be identical
for all bands. The best-fit results, calculated by minimizing the sum of $\chi^2$ 
from the S, M and H bands, are listed in Table 2.
The sum of $\chi^2$/$dof$ for the S, M and H bands was 
597.5/540 for the sharply-broken model and
599.1/540 for the slowly-bending model.
The summed values of $\chi^2$/$dof$ when $f_{\rm b}$ was allowed to vary in 
photon energy ($\gamma$ and $\beta$ frozen)
was 584.9/540 (sharply-broken) or 588.3/540 (slowly-bending). 
We conclude that a simple broken power-law model where $f_{\rm b}$ does not vary with energy cannot accurately
describe the S, M and H-band PSDs of Mkn 766.

However, there is an alternate model that fulfills the requirements of having 
$f_{\rm b}$ energy-independent and providing a good fit to the S, M and H-band PSDs.
This is a sharply-broken power-law model with the addition of a Lorentzian
component whose strength increases in photon energy:
\[P(f)= \left\{ \begin{array}{ll}
           A(f/f_{\rm b})^{-\gamma} + P_{\rm L} + P_{\rm Psn},  & f \le f_b \\
           A(f/f_{\rm b})^{-\beta} +  P_{\rm L} + P_{\rm Psn},   & f > f_b \end{array}  
\right.  \]
where
\[ P_{\rm L} = \frac{ R^2 f_{\rm L} Q/\pi}{ f_{\rm L}^2 + Q^2(f - f_{\rm L})^2}  
 \] 
$f_{\rm L}$ is the Lorentzian centroid frequency. $Q$ is the coherence or 
quality factor. $R$ is the normalization factor of the Lorentzian; for high 
$Q$ values, $R$ is the fractional rms variability amplitude of the Lorentzian.
For values of $Q < 2$, the Lorentzian is very broad and
cannot be easily distinguished from band-limited or broadband noise,
e.g., the broken power-law component. Lorentzians with
higher $Q$ values are called QPOs (e.g., van der Klis 2006).

The addition of the Lorentzian component to the sharply-broken power-law
model complicates the model-fitting process. For that reason, in order for
the model-fitting results to be reasonable and of some use, we decided to
make the following plausible assumptions regarding the parameter values of
the Lorentzian component, based mainly on previous results from similar
model fits to the XRBs PSDs: First,  we are interested in fitting a  
peak-like component, and so we only consider a Lorentzian with 
$Q \gtrsim 3$ in order to avoid fitting a component that resembles additional broadband 
noise. We cannot place tight constraints on $Q$ but have 
found that values of $\sim 4-5$ were the most plausible in terms of providing a
good fit to the PSDs. For that reason, all model fits were performed 
with Q kept fixed at 5. Second, we forced $f_{\rm L}$ to be the same in all three PSDs 
(i.e., we assumed that $f_{\rm L}$ is intrinsically energy-independent) and we 
further constrained its value to lie in the range $ 10^{-3.22} - 10^{-3.50}$ Hz;  $f_{\rm L}$ 
was tested in increments of 0.02 in the log. Finally, $R$ was left as a free parameter, tested in increments of 0.5$\%$.
The break frequency $f_{\rm b}$ of the broken power-law 
continuum component was not constrained to be equal for all bands, but was
constrained to be $\leq 10^{-3.60}$ Hz. Finally, we left $\beta$ as a free
parameter but kept $\gamma$ fixed at $-1.46$ for simplicity.

Table 3 lists the best-fit results obtained
from finding the minimum value of $\chi^2$ summed over the S, M and H bands.
The data, best fit models, and residuals are plotted in Figure 6.
First, we note that both $f_{\rm b}$ and
the power-law slope above the break $\beta$ are similar for all bands. 
Second, we find that
the best-fit log($f_{\rm L}$ (Hz)) is $-3.34 \pm 0.04$  ($4.6\pm0.4 \times 10^{-4}$ Hz). 
Finally, we note that $R$ increases with photon energy:
it is formally consistent with an upper limit of only 1.5$\%$ in the S band, but is
$5.5\pm0.5 \%$ in the H band. This is also demonstrated by Figure 7, 
which shows derived confidence contours of $R$ versus $f_{\rm L}$ for each band separately.

Using a slowly-bending model to fit the continuum yielded
virtually identical values of $\chi^2$, $R$, $f_{\rm L}$
and identical uncertainties on $R$ and $f_{\rm L}$
in each band; the best-fit value of $f_{\rm L}$
was only 0.02 lower in the log compared to using
the sharply-broken model.

The temporal frequency resolution of this PSD, despite being the
highest for any binned AGN PSD to date,
is not high enough to break model degeneracies and
simultaneously constrain the curvature of the break and the profile
of the Lorentzian component, especially since the Lorentzian
is located so close to the break frequency.
In addition, the broken power-law model and the model with the 
Lorentzian component both yield similarly good fits to the PSD, so we
cannot exclude one or the other based on the quality of fit alone.

Nonetheless, the PSD is of sufficient quality to detect energy-dependent
behavior that has not been observed previously in an AGN.
Either $f_{\rm b}$ increases with photon energy (in which case the 
broadband PSD shape is likely the same for all energies),
or $f_{\rm b}$ and $\beta$ are energy-independent and there exists 
a peak-like component, which we model as a Lorentzian, 
near $ 5 \times 10^{-4}$ Hz whose strength increases with photon energy.

\section{Cross-spectral Properties}

The cross-spectrum is used to quantify interband correlations
as a function of temporal frequency; analogous to the fact that
the PSD is the Fourier transform of the ACF, the cross-spectrum
is the Fourier transform of the CCF. The cross-spectrum is a complex
number;  coherence and time lags can be derived from
the squared magnitude and argument, respectively.
Because these quantities in compact accreting objects
tend to be temporal-frequency dependent, 
they cannot be studied using cross-correlation functions alone.
Further details of the cross-spectrum and its relevance to
XRB and AGN observations can be found
in e.g., Vaughan \& Nowak (1997), Nowak \et\ (1999a), 
Papadakis, Nandra \& Kazanas (2001) and VFN03.

\subsection{The Coherence Function}

The coherence function $\gamma^2(f)$
between two time series measures the fraction of 
mean-squared variability of one series that can be attributed 
to other, as a function of temporal frequency. If the two time
series are related by a simple, linear transfer function, then 
they will have unity coherence. The coherence is defined as the 
magnitude squared of the cross spectrum normalized by the product 
of each light curve's PSD. Here, we use the discrete versions 
of the cross-spectrum and coherence functions, given in e.g.,
$\S$5.2.2 of VFN03. 

We correct the coherence function for the influence of Poisson noise
by following Eq.\ 8 of Vaughan \& Nowak (1997).
We assume that the condition of high coherence and high powers
required to use this formulation are valid. Specifically,
the intrinsic variability power must be at least a factor of 
a few times  $\sqrt{m} \times P_{{\rm Psn}}$,
where $m$ is the coherence binning factor (quantified below);
this condition is thus valid for temporal frequencies up to roughly
$ 1-2 \times 10^{-3}$ Hz.

Coherence between the S--M, S--H and M--H band pairs
was measured separately for each of the eight light curves,
then combined, sorted 
in temporal frequency, and binned in groups of $m$=30 to form the 
final coherence function estimate.
As shown in Figure 8, the coherence is near unity 
($\gtrsim 0.8-0.9$) over the lowest temporal 
frequencies sampled for the S--M and M--H pairs. 
For S--H, the coherence is a bit lower, $\sim 0.6-0.8$,
at these temporal frequencies.
At the highest temporal frequencies sampled, 
the coherence significantly drops sharply towards zero in all cases.

This cutoff at high temporal frequencies
can be intrinsic, but could also be due to Poisson noise.
Monte Carlo simulations were used as a check on the accuracy
of the coherence estimates and to determine if significant
deviations from unity could be attributed to the effect of 
Poisson noise. For the 2005 long-look, two light curves
of duration several Msec were simulated using the
same random number seed and assuming identical PSD shapes 
(the average of the best-fitting singly-broken PSD
model for each band pair) to produce two light curves 
with intrinsic unity coherence. The simulated light curves were
rebinned and resampled to match the
observed light curves (e.g, broken into 
seven smaller light curves matching the sampling
of the observed ones), and rescaled to match the mean observed
count rates in the softer and harder bands.
Poisson noise was then added to the data according 
to the Poisson distribution, i.e., data were randomly deviated using 
a Gaussian convolved with the square root of the observed mean error 
squared divided by the sampling time.
Simulated light curves for the 2001 data were produced in the same way.
One hundred sets of simulated data were produced. For each set, 
the coherence and its uncertainty was calculated as for the real data
(e.g., combining the eight simulated light curves).
At most temporal frequencies below $\sim 10^{-3}$ Hz,
the simulated coherence was reasonably close to unity, and 
the uncertainty in the coherence was in reasonable agreement
with the scatter, indicating that the coherence estimation was 
reasonably accurate. The 90$\%$ and 95$\%$ lower limits to the coherence 
were determined. Any measurement of the observed coherence  
function lying below these lines thus indicates a drop in coherence 
that is intrinsic to the source, and not an artifact of the Poisson 
noise, at 90$\%$ or 95$\%$ confidence. For most temporal frequencies
below $\sim 10^{-3.5}$ Hz, the artificial drop in coherence is small.

The energy-dependence of the high-frequency PSD slope
could also contribute to a reduction in measured coherence.
The Monte Carlo simulations were repeated, assuming
the best-fit sharply-broken power-law model shapes. 
The effects of Poisson noise
were again added. The simulated coherence was
typically only $\sim 1\%$, $3-4\%$ and $1-3\%$
lower for the S--M, S--H and M--H bands, respectively,
compared to simulations assuming identical PSD
shapes. The effect of having differing PSD
shapes is thus small. The 90$\%$ and 95$\%$ lower limits to the coherence 
were determined and are plotted in Figure 8.
The observed deviations from
unity are generally larger than the artificial drop due to Poisson noise. 
The decrease in intrinsic coherence as the energy separation of the bands
increases and as temporal frequency increases towards
$\sim 10^{-3}$ Hz is also likely intrinsic. 

Having quantified the decrease in coherence e.g., due to
Poisson Noise at each temporal frequency, we added the mean value of the
drop in coherence from the Monte Carlo simulations to the observed coherence
to obtain an estimate of the 'intrinsic' coherence, plotted in the bottom panel of
Figure 8. We fit the 'intrinsic' coherence functions with
exponential decay functions of the form 
$\gamma^2(f) = C$exp$(-(f/f_{\rm c})^2)$, where
$f_{\rm c}$ is the temporal frequency cutoff in Hz
and $C$ is a constant,
ignoring points above $1.2 \times 10^{-3}$ Hz, and 
using least-squares fitting, assuming equal weighting to all points.
We find that log($f_{\rm c}$) for the S--M, S--H and M--H
coherence functions are $ -2.82 \pm 0.03$, $-2.80 \pm 0.03$
and $-2.81 \pm 0.03$, respectively.   %%%%%  (errors correspond to $\Delta\chi^2$ = 2.71).

\subsection{Temporal Frequency-Dependent Time Lags}

Cross-correlation functions (CCFs) for XRBs and AGN
generally show a peak near zero lag
but are often asymmetrically skewed towards hard lags 
(e.g., M$^{\rm c}$Hardy \et\ 2004), suggesting there is a soft-to-hard time lag of 
some sort, but it is not a simple delay. 
We measured the CCFs for the 2005 data
using e.g., the Interpolated Correlation Function (White \& Peterson 1994). 
Because large gaps such as those
between revolutions complicate analysis, we calculated CCFs
for each individual revolution separately.
As expected, all peak lags are consistent with zero
and CCFs exhibit a slight soft-to-hard asymmetric skew. 
The logical direction is to explore temporal frequency-dependent time lags via the
cross-spectrum. 

Obtaining sensible measurements of time lags via the cross-spectrum 
is difficult for observations of AGN, however; it requires the 
combination of a large quantity of uninterrupted data with a high 
sampling time resolution and a high variability-to-Poisson noise
ratio. VF03 did not present lags for the 2001 data; we also attempted 
to derive time lags from the 2001 pn data and did not detect any obvious 
time lag trends, likely due to the combination of intrinsically small 
lag values (see below), high scatter, and low temporal frequency 
resolution.

We used Eqs.~28 and 29 in $\S$5.3.1 of VFN03 to calculate the phase lags
$\phi(f)$; time lags 
were then obtained as $\tau(f) = \phi(f)/2\pi$$f$.
Time lags were calculated for each of the eight light curves
separately, then combined, sorted in frequency, and binned in groups of
50. Errors on phase lags were calculated using Eq.~16 of Nowak \et\ (1999a);
in this formulation we used the coherence function appropriate for
each binned point. Sensible lags (e.g., lags all have the same sign and 
show a trend) were obtained for the 4 lowest
temporal frequency bins, spanning 
$ 4.4 \times 10^{-5} - 3.3 \times 10^{-4}$ Hz.
The results are shown in Figure 9.
Errors are large, primarily because coherence
in this temporal frequency range is not always exactly unity.
Using Eq.~16 in Nowak \et\ (1999a), we estimated the effective 
sensitivity limit on time lag detections
due to Poisson noise, using the 
PSD shape and normalization 
from best-fit M-band singly-broken
PSD model shape in $\S$4.
Limits are shown as dotted lines in Figure 9.

Monte Carlo simulations, based on the simulated light curves
used for the coherence functions, were performed to verify that
the time lags were not an artifact of Poisson noise nor of
the differing PSD slopes between two bands. 
For all simulations, time lags were $<$24 s for the two lowest 
temporal frequency bins and $<$15 s for the next two bins. 
We can reasonably conclude that the observed time lags are not an artifact
of Poisson noise or PSD shape differences, and are
henceforth assumed intrinsic to the source.

The time lag spectrum displays the general qualities seen in
many previous time lag spectra: a general increase towards lower 
temporal frequencies and as the energy separation of the bands increases.
Assuming that the time lags are proportional
to $\tau(f) = C f^{-1}$ for simplicity
allows us to calculate fractional time lags, i.e.,
the lags' fraction of the temporal frequency.
We find that $C$ = 0.002, 0.009, and
0.007 for S--M, S--H and M--H, respectively.
These values of $C$ are roughly correlated with
values of log($E_2$/$E_1$)  = 0.32, 1.03 and 0.72, respectively,
where $E_1$ and $E_2$ are the relatively softer and harder 
average photon energies in each band ($\S$4.2.1).
Similar correlations have been reported previously
in XRBs and in the NL Sy1 Ark 564 (Ar\'{e}valo \et\ 2006a).
The physical implications of these time lags and comparisons to other
objects will be discussed further in $\S$6.

\section{Discussion}

An \xmm\ observation of Mkn 766 spanning six revolutions in 2005
caught the source in a relatively low flux state;
with the average 0.2--12 keV flux about half of the flux
during the 107 ksec \xmm\ observation in 2001.
Reduction of the 2005 observation yielded seven long, virtually uninterrupted
light curves of this highly-variable source totaling 442 ksec of good 
exposure time.

We find no evidence for the underlying variability process to 
have changed from 2001 to 2005, despite a factor of 2 difference 
in average flux. Specifically, the PSD shapes and coherence functions
between 2001 and 2005 are consistent 
and both observations appear to have consistent linear
rms-flux relations and constant $<$$\fv$$>$ values. 
The Mkn 766 light curves are thus stationary over a four year duration 
as far as the second-order moments are concerned.

Combining the 2005 and 2001 observations, we have constructed a binned
PSD with the highest temporal frequency resolution in log space
for an AGN PSD to date. We find energy-dependent behavior not
reported previously in AGN PSDs, as discussed in $\S$6.1.
In addition, time lags have been detected in Mkn 766 for the first time;
$\S$6.2 discusses the cross-spectral results and compares
them to those of other compact accreting objects.
Finally, the results are discussed in the context of phenomenological 
variability models in $\S$6.3.

\subsection{Summary of PSD Results}
  
The high temporal frequency resolution of the PSDs allows us to place tight 
constraints on localized features (a PSD break or the presence of a peak-like
feature, i.e., a QPO) in any model we fit. Our initial approaches to 
model the PSD centered on broken power models. Model fits cannot formally 
distinguish between a sharp break or a slow rollover, despite the high 
frequency resolution; both models fit the data well. In general, there are no 
obvious trends in the power-law slope above or below the break, suggesting the 
existence of a ``universal'' PSD shape at all bands whose break scales with 
photon energy. In that case, fit parameters may depend on the temporal frequency 
range being studied. Our main result is that $f_{\rm b}$ increases with photon 
energy. Neither the increase in $f_{\rm b}$ nor an energy-independent PSD
shape has been conclusively demonstrated previously in AGN.

\subsubsection{The increase in $f_{\rm b}$ with photon energy}

Our estimate for the T-band break frequency is
log($f_{\rm b}) = -3.36\pm0.04$ (sharply-broken model, all parameters free).
This value is consistent with that of VF03, but 
the uncertainty is much lower compared to VF03 or even to VFN03 for
the three \xmm\ revolutions of data for MCG-6-30-15.
For the first time in an AGN PSD, the break frequency is 
seen to increase with energy, a result that is 
robust to the specific broken power-law model shape used.
The break frequency tends to increase by 
$\sim 0.2-0.4$ in the log between the S and H bands (a factor of
$\sim$10 in average photon energy).

Most previous Seyfert PSDs simply did not have sufficient
temporal frequency resolution to detect 
significant changes in $f_{\rm b}$ with photon energy such as those seen here.
It is plausible that for a limited dynamic range PSD with low frequency resolution, the 
effect of having $f_{\rm b}$ intrinsically increasing with photon energy
could lead instead to apparent measurements of the power-law
slope above the break, $\beta$, flattening with energy.
Indeed, when we fit the Mkn 766 PSD with $\gamma$ and $f_{\rm b}$
frozen, $\beta$ flattens with energy, %%%particularly from the M to H bands,
suggesting that previously observed trends of $\beta$ with photon energy
may instead be due to an intrinsic energy dependence of $f_{\rm b}$. 
Alternatively, it could be the case that this phenomenon
might exist only in Mkn 766;
high-quality, high-resolution PSDs are needed for a sample of Seyferts
to determine whether this is the case.

%%%%%% last paragraph of 6.1.1:

The fact that the observed increase in $f_{\rm b}$ with energy is small
suggests that PSD measurement spanning a range of   
photon energies wider than those used here 
would be needed to further separate $f_{\rm b}$ in energy.
In the case of XRBs, it is possible that 
this phenomenon may not have been 
reported due to insufficient spread in the hard X-ray photon energies studied.

\subsubsection {The possible presence of a Lorentzian component}  

The alternate to modeling the PSD with $f_{\rm b}$ as a free parameter
is to assume that $f_{\rm b}$ is energy-independent. Achieving a good fit 
requires the presence of a peak-like component, which we model as a
Lorentzian with $Q=5$, that is centered at $4.6\pm0.4 \times 10^{-4}$ Hz and whose 
rms variability $R$ increases from $< 1.5\%$ in the soft band to $5.5\pm0.5 \%$ 
in the hard band. Such a detection, though model-dependent, is also a new result for AGN.

Following the definition of van der Klis (2006), the Lorentzian component
can be identified as a QPO. Most previous reports of deterministic behavior 
or claims of QPOs in AGN have been refuted, have failed the test of repeated 
observations, or did not use the appropriate statistical significance tests 
(see Vaughan \& Uttley 2005 and Vaughan 2005).
We our basing our claim on modeling of a PSD that is properly binned,
has well-defined errors, has very high temporal frequency resolution,
and is derived from a large quantity of high-quality data.
We sample almost 300 cycles of this feature throughout the duration.

We reiterate that our claim of the presence of a QPO in the PSD
is highly model-dependent, since a simple broken power-law model can fit the data well.
The requirement for the QPO arises only when $f_{\rm b}$ is assumed to be 
energy-independent. Both models adequately describe the energy dependence of the PSD well, though
Occam's Razor may have one prefer
the simple, broken-power law model, since it consists of only the broadband PSD component
and does not make any {\it a priori} assumption about the behavior of $f_{\rm b}$ with energy.

Assuming that the QPO at $ 4.6 \times 10^{-4}$ Hz 
is real, there are intriguing similarities 
to those observed in XRBs. QPOs whose rms variability $R$ increases with photon energy
have been observed in XRBs. For example, the 450 Hz QPO in GRO J1655--40 and
the 40 Hz QPO in GRS 1915+105 are detected only above 13 keV
by Strohmayer \et\ (2001a) and (2001b), respectively.  
In Mkn 766, the value of $R$ is highly model-dependent, but also
increases with photon energy and displays roughly 
similar values, from $< 1.5\%$ at $\sim$0.4 keV to 
5.5$\%$ at $\sim$4 keV. It is not unreasonable to compare 
the (relatively softer) energy bands in Mkn 766 with the (harder)
energy bands in XRBs, since softer bands in XRBs may be dominated by thermal disk emission.

Photon indices in the low/hard, high/soft and very high state in XRBs tend to be $\sim 1.5-2.0$,
$\gtrsim 2-3$, and $\sim 2.4-2.9$, respectively (e.g., 
McClintock \& Remillard 2006; Homan \et\ 2005). The photon 
index of the power-law continuum
in Mkn 766 is $\sim 2.3-2.4$ (e.g., Turner \et\ 2006, Miller \et\ 2006).
A direct comparison between Mkn 766 and
XRBs in the very high state therefore might not be rejected on those grounds.
QPOs, however, tend to appear only in XRB states when the thermal disk 
emission does not dominate the energy spectrum
(van der Klis 2006, McClintock \& Remillard 2006). 
The existence of a QPO in Mkn 766 would therefore argue against
Mkn 766 being an analog of the (thermal-dominated) high/soft state in XRBs.

One may ask whether the possible QPO in Mkn 766 
is the analog of low-frequency (LF) or high-frequency (HF) QPOs. 
Type A and B LF QPOs usually appear at 4--9 Hz. Assuming that QPO 
frequencies scale linearly with black hole mass, and assuming
10 $\Msun$ for the mass of stellar-mass black holes, the mass of the
black hole in Mkn 766 is thus estimated to be $1-2 \times 10^5 \Msun$.
This is somewhat low compared to estimates of Mkn 766's black hole mass:
Botte et al.\ (2005) measure a stellar velocity dispersion 
$\sigma_{*}$ of 81 $\pm$ 17 km s$^{-1}$ in Mkn 766. Using the relation 
between $\sigma_{*}$ and $M_{\rm BH}$ as parameterized by 
Tremaine et al.\ (2002), this suggests $M_{\rm BH} = 
3.6^{+3.5}_{-1.9} \times 10^{6} \Msun$. Estimates 
using the width of the H$\beta$ line and the Kaspi \et\ (2000) relation
lie in the range $ 0.8 - 10.0 \times 10^6 \Msun$ (Wandel 2002; 
Bian \& Zhao, 2003a, 2003b, 2004; Botte \et\ 2005). On the other hand,
that fact that XRB QPOs tend to appear only in states when the thermal disk 
emission does not dominate the energy spectrum suggests that QPOs
are associated with the corona instead of the disk (Reig \et\ 2006; 
van der Klis 2006, McClintock \& Remillard 2006). QPO frequencies may thus
scale with some other property besides or in addition to black hole mass. 
Type C LF QPOs tend to have $R$ values which may be too high to compare to Mkn 766.

HF QPOs usually are seen near 100--300 Hz, and usually with $R \sim 1-3\%$.
They are often found in pairs, at frequencies of 2 and 3 times an unseen
``fundamental'' frequency $f_0$. The Mkn 766 PSD does not have the 
resolution to determine if the QPO corresponds to the 
2$f_0$ or the 3$f_0$ peak being dominant, or if it is a blend of the two.
Based on observations of three black hole XRBs, McClintock \& Remillard 
(2006) have suggested that $f_0$ scales linearly with black hole mass
$M_{\rm BH}$ as 931 Hz ($M_{\rm BH}/\Msun)^{-1}$, assuming similar 
values of the spin parameter; this linear scaling suggests that HF QPOs 
in black holes may represent disk oscillations described by general 
relativity (see Abramowicz \& Kluzniak 2001). Extrapolating this 
relation to Mkn 766 and assuming $f_0 = 1.8-2.4 \times 10^{-4}$ Hz,
this implies a black hole mass of $ 3.9-5.2 \times 10^6 \Msun$,
consistent with the optically-estimated black hole masses. 
On the other hand, HF QPOs are not always found very close to 
the break in the broadband component, as is the case here. 

In the end, the community needs very high-quality data on a reasonably sized sampled
of AGN to be able to confirm or refute the presence of 
QPOs in AGN in order to shore up the comparisons between 
AGN and XRBs. The Mkn 766 observations are an important step towards this goal,
demonstrating that high-quality data
in the form of a collection of long, uninterrupted exposures 
can successfully yield an enormous payoff for AGN variability investigations. 
%%%%%%% and demonstrating links between AGN and XRBs.

\subsection{Summary of cross-spectral results and comparison to other objects}

The coherence is generally flat at the lowest temporal frequencies 
probed, with values typically $\sim 0.8-0.9$ 
for S--M and M--H and
$\sim 0.6-0.8$ for S--H. The coherence drops 
towards zero near $\sim 10^{-3}$ Hz; Monte Carlo simulations 
prove that these deviations from unity are
intrinsic to the source and not the result of
Poisson noise or energy-dependent PSD shapes.
The decrease in coherence as energy separation increases has been
reported previously for Mkn 766 by VF03; we in fact see values of the coherence
similar to VF03.
This similarity again supports the conclusion
of $\S$3, that there is no evidence for
the underlying variability process to have changed
significantly between 2001 and 2005 despite the overall
lower average flux in 2005.

In other black hole systems, the coherence at temporal frequencies
below the break is generally between 0.8 and 1.0,
though coherence between bands similar to the S and H bands used here
can be as low as $\sim 0.6-0.7$ (e.g., VFN03 \& M$^{\rm c}$Hardy \et\ 2004);
Mkn 766's coherence functions thus agree with those in
other compact objects.
As explained in Vaughan \& Nowak (1997), coherence between two energy bands can be lost if
the transfer function relating the light curves is non-linear, or if there are multiple uncorrelated flaring 
regions.  The fact that coherence in Mkn 766 and several other compact objects
is not exactly unity may support some degree of non-linearity present
in the transfer functions relating relatively softer and harder bands,
and which increases for larger energy separations.

%  Coh below break:
%        Ark564: 0.9-1.0   for M-H   
%        N4051   0.8-1.0 for S-M
%         n4051    0.6-1.0 for S-H
%        MCG6      S-M and M-h: 0.8-1.0 
%        mcg6          S-H: 0.7-0.9
%        Cyg X-1   Always 0.9 - 1.0 (Nowak 99) 

The coherence function between the S and H bands 
drops to $\sim$1/$e$ around $1 \times 10^{-3}$ Hz 
in both MCG--6-30-15 (VFN03) and Mkn 766, while
the drop in NGC 4051's S-H coherence occurs at roughly
$\sim 4 \times 10^{-3}$ Hz (M$^{\rm c}$Hardy \et\ 2004).
MCG--6-30-15's black hole mass is roughly 
$3 \times 10^{6} \Msun$ (M$^{\rm c}$Hardy \et\ 2005), 
similar to that for Mkn 766,
while that for NGC 4051 is $1.9\pm0.8 \times 10^6 \Msun$ (Peterson \et\ 2004). 
A larger sample of coherence functions of Seyferts spanning a larger black hole
mass range is needed to verify if coherence time scales scale with 
black hole mass, just as PSD break time scales do.

As explained by e.g., Nowak \& Vaughan (1996) and Nowak \et\ (1999a),
the sharp drop in coherence at temporal frequencies above
the PSD break could indicate that the corona is dynamic on such short time scales.
Such short time scales could be associated with formation time scales or 
changes in temperature and/or the physical structure of the corona.
The relatively high levels of coherence at low temporal frequencies
could indicate that the corona is effectively static on long time scales.

% exponential decay     SM,SH,MH  for mkn 766   -2.70,-2.83,-2.72         C(0.4) = 8,7,9 e-4 Hz 
%                                     mcg6    8e-4                        c(0.4) = 
%                                      n4051                               c(0.4): MH: <<1e-2    SH: 400 sec 

%%%%%%%%%%

We measure temporal-frequency dependent phase lags in Mkn 766 for the first time; lags increase
towards lower temp frequency and as energy separation of bands increases,
as seen in other black hole systems.
Fractional time lags are about 0.01--0.02 the Fourier time scale.
These values are consistent with most values measured
so far for other AGN and many XRBs (see e.g., Figure 8 of Ar\'{e}valo \et\ 2006a), with
fractional time lags usually measured to be
$\lesssim 0.2$, though Ark 564's time lag spectrum (Ar\'{e}valo \et\ 2006a)
seems to be an exception among AGN. 

Assuming a PSD break frequency of 
$10^{-3.4}$ Hz, the Mkn 766 lag spectrum is
probing temporal frequencies $0.14-1.00$ times the
break frequency. We can compare Mkn 766's fractional
lags to those measured for Cyg X-1 within $\pm$ a decade of temporal 
frequency of the PSD breaks.
In the low/hard and high/soft
state of Cyg X-1, fractional time lags tend towards 0.03--0.1 (e.g., Nowak \et\ 1999a),
while in the intermediate state, 
fractional time lags are higher, $\sim 0.1-0.2$ (e.g., Pottschmidt \et\ 2000).
On the basis of these fractional lag values, Mkn 766 is likely not 
a supermassive analog of Cyg X-1 in the
intermediate-state, though we cannot determine
on the basis of lags alone if analogy with the low/hard
or high/soft state is appropriate.
Furthermore, the time lag spectrum is not of high
enough quality to distinguish between an intrinsic lag
spectrum which is step-like, as in the 
intermediate state of Cyg X-1, or adheres to a form
more resembling a power-law.

%%%%%%%%%%%%%%%%%%%%%%%%%%%%%%%%%%  6.3 Physical models

\subsection{Phenomenological Variability Models}

Many models have been invoked to explain the red noise variability 
properties of Seyferts and XRBs, including shot-noise models (e.g., 
Merloni \& Fabian 2001), rotating hot-spots on the surface of the
accretion disk (Bao \& Abramowicz 1996), self-organized criticality 
(``pulse avalanche'' models; Mineshige, Ouchi \& Nishimori 1994), 
magnetohydrodynamical instabilities in the disk (Hawley \& Krolik 2001), 
and inwardly-propagating fluctuations in the local accretion rate 
(Lyubarskii 1997).  As noted by
Uttley, M$^{\rm c}$Hardy \& Vaughan (2005),
shot-noise models have difficulty reproducing the linear rms-flux relation,
and, as noted by Vaughan \& Nowak (1997),
they lead to low observed coherence values unless each separate 
emitting region independently has the same linear transfer function 
between energy bands.

Thermal Comptonization of soft seed photons by a hot corona is a 
likely explanation for the X-ray emission. It is supported 
by X-ray energy spectra, and 
the soft-to-hard time lags could be attributed to the
time scale for seed photons to diffuse through the corona and undergo 
multiple up-scatterings. However, the simplest Comptonization
models do not predict time lags that depend on temporal
frequency, as observed.

%%%%  more scatterings, will have rapid variability washed out, e.g., variability amplitudes at 
%%%   time scales shorter than the PSD break time scale should decrease with increasing energy, 
%%%   contrary  to observations.
%%%%However, the energy dependence of the PSD could be obtained with a corona whose
%%%%temperature increased towards smaller radii (Kazanas, Hua \& Titarchuk 1997).

A model in which inwardly-propagating variations in the 
local mass accretion rate $\dot{m}$ are responsible for the observed X-ray 
variability seems to be able to explain many of the observational 
results in Seyferts and XRBs (Lyubarskii 1997, Churazov, Gilfanov 
\& Revnivtsev 2001), Kotov \et\ 2001, Ar\'{e}valo \& Uttley 2006;
see also VFN03 and M$^{\rm c}$Hardy \et\ 2004). 
In this model, variations at a 
given radius are associated with the local viscous time 
scale, so relatively smaller radii are associated with relatively 
more rapid variations. Perturbations in $\dot{m}$ 
propagate inward, until they reach, and modify the emission of,
the central X-ray emitting region. The resulting
PSD has a 1/$f$ form across a broad range of temporal frequencies, 
the net sum of variations from a wide range of radii.
However, if the X-ray emitting region is extended,
it essentially acts like a low-pass filter on these
1/$f$ variations: 
with relatively higher-frequency variations imprinted
on the light curve originating only from the smallest
radii, this produces a cutoff in the high-frequency PSD,
as is commonly observed.
Kotov \et\ (2001) suggested that the spectrum can be a function of radius, 
with relatively harder emission emanating from smaller radii, and 
therefore associated with more rapid variability.
This class of models thus predicts several additional
properties which are observed in Seyferts and XRBs, namely
the presence of the linear rms-flux relation, and
the dependence of coherence and time lags on
temporal frequency and on the energy separation of the bands.  

Ar\'{e}valo \& Uttley (2006) show that
if relatively harder X-ray bands
are associated with emissivity profiles that are
more centrally concentrated,
and if the PSD break is at least
partly due to filtering produced by having an extended
X-ray emission region,  then the 
harder bands will retain more variability power at the
highest temporal frequencies, leading to  
break frequencies in the broadband PSD
which increase with increasing photon energy.
We note that the PSDs of Mkn 766 as shown in Figure 5b,
bear a striking qualitative resemblance to the simulated PSDs 
in Figure 2 of Ar\'{e}valo \& Uttley (2006).
The degree of curvature in the Mkn 766 PSDs is 
uncertain, and rollover frequencies are relatively close together,
so it is not straightforward to relate break frequencies
to emissivity index or profile. However, we note that
the 'soft' and 'hard' PSDs in Figure 2 of
Ar\'{e}valo \& Uttley (2006), simulated assuming
emissivity indices which differ by 2,
yield break frequencies which are separated by about a decade.
In Mkn 766, the S- and H-band break frequencies
are not as separated, suggesting that, in the
context of this model, the emissivity indices for
the S- and H- bands differ by less than 2.
The relatively small fractional time lags observed in
Mkn 766 (0.01--0.02 for the S--H bands) 
are consistent with a difference in emissivity indices of $\sim$0.5 to a few.

%%% {\bf \footnotesize (I got the impression from VF03 that 
%%% they though coherence, especially S--H, was a bit lower
%%% compared to other objects. I'm not sure I agree.
%%% If it is true, though, does that mean that
%%% damping could play more of a role in Mkn 766's
%%% disk compared to other objects?) }

\section{Conclusions}

We have analyzed the energy-dependent variability properties
of the Narrow Line Seyfert 1 Mkn 766 using 
light curves obtained from two {\it XMM-Newton} observations.
An observation spanning six revolutions
in 2005 yielded seven uninterrupted EPIC light curves
totaling 442 ksec of good exposure time; an observation in 2001 yielded
an eighth uninterrupted light curve with an exposure of 107 ksec.
A linear rms-flux relation, or equivalently, constant $<$$\fv$$>$,
was reported for the 2001 observation by Vaughan \et\ (2003a).
These relations are confirmed for the 2005 observation,
and shown to be consistent with the corresponding relations from the
2001 observation. Comparing the 2001 and 2005 PSDs
shows the PSDs to be consistent within the expected scatter.
Variability properties between the 2005 and 2001 observations 
are thus consistent with each other, despite a 0.2--12 keV flux
in 2005 which is about half that of the 2001 observation.
We combined the 2001 and 2005 data sets in order to
probe power- and cross-spectral properties with temporal frequency
resolution that is unprecedented for an AGN. 

We have explored two parameterizations of the energy-dependent PSD.
Both fit the data equally well, but in either case,
they signify behavior never seen previously in an AGN PSD.
Fitting simple broken power-laws, 
we find PSD breaks at nearly the same temporal frequencies as VF03,
but the break frequencies $f_{\rm b}$ are seen to increase with
photon energy, typically by $\sim$0.2--0.3 in the log between $\sim$0.4 keV
and $\sim$4 keV emission. There is no obvious trend in PSD power-law slope, 
and the data are consistent with a universal
PSD shape whose break frequency increases with photon energy.
Previous Seyfert PSDs lacked the temporal frequency resolution
to cleanly detect the energy dependence in $f_{\rm b}$.
Keeping $f_{\rm b}$ fixed with energy does not provide a good description of
the energy-dependent PSD. 
However, we can achieve a good fit using a sharply-broken power-law, where
the break frequency and power-law slope above the break are roughly energy-independent,
but we must include a narrow peak component, i.e., a QPO, 
at $4.6\pm0.4 \times 10^{-4}$ Hz.
Interestingly, the rms variability amplitude of this QPO increases 
with photon energy in a manner similar to QPOs seen in the PSDs of black hole XRBs.

Cross-spectral properties are qualitatively consistent with those measured for
black hole systems previously.
The coherence is generally flat at the lowest temporal frequencies probed, with
values typically $\sim 0.6-0.9$. The coherence drops 
towards zero at temporal frequencies higher than roughly 10$^{-2.8}$ Hz; 
Monte Carlo simulations
indicate that the drop is intrinsic to the source and not an artifact
of Poisson noise or differing PSD shapes.

Temporal frequency-dependent time lags are measured from the cross-spectrum for
the first time in Mkn 766. As is the case with time lags in other
objects, lags increase as temporal frequency increase and as energy separation
of the light curves increases.
Fractional time lags are typically $\lesssim$0.01, consistent with
many other Seyferts as well as with the high/soft and low/hard
states of Cyg X-1. 

The results were discussed in the context of several variability models,
including models incorporating inwardly-propagating fluctuations
in the local mass accretion rate, e.g., Lyubarskii (1997).
Notably, the observed increase in the PSD break frequency with photon energy
is qualitatively consistent with the prediction 
put forth by e.g., Ar\'{e}valo \& Uttley (2006), who model variations
in a disk with relatively harder X-ray emission being more
centrally concentrated.

\acknowledgements
We thank the referee for providing useful comments.
This work has made use of observations obtained with {\it XMM-Newton}, an 
ESA science mission with instruments and contributions directly 
funded by ESA member states and the US (NASA).

%%%%%%%%%%%%%%%%%%%%%%%%%%%%%%%%%%%%%%%%%%%%%%%%%%%%%%%%%%%%%%%%%%%%%%%%%%%%%%%%%%%%%

%%%\clearpage

%%%%%%%%%%%%%%%%%%%%%%%%%%%%%%%%%%%%%%%%%%%%%%%% TABLES %%%%%%%%%%%%%%%

\begin{deluxetable}{lccc}
\tabletypesize{\footnotesize}
\tablewidth{6.5in}
\tablenum{1}
\tablecaption{Unbroken Power-law Model Fits to the PSD \label{tab1}}
\tablehead{
\colhead{Band} & \colhead{$\chi^2$/$dof$} & \colhead{$\beta$} & \colhead{$A$}  }
\startdata
T & 299.3/180 & 2.34$\pm$0.06          & 0.99$\pm$0.03 \\
S & 137.4/180 & 2.14$\pm$0.08          & 0.79$\pm$0.04 \\
M & 258.3/180 & 2.22$\pm$0.06          & 1.00$\pm$0.03   \\    
H & 278.4/180 & 2.14$\pm$0.08          & 1.16$\pm$0.03 \\
\enddata
\tablecomments{$A$ is the PSD amplitude in units of Hz$^{-1}$ at $f_0$ = 10$^{-3.50}$ Hz.}
\end{deluxetable}

\begin{deluxetable}{lccccc}
\tabletypesize{\footnotesize}
\tablewidth{6.5in}
\tablenum{2}
\tablecaption{Sharply- and Slowly Broken Power-law Model Fits to the PSDs \label{tab2}}
\tablehead{
\colhead{Band} & \colhead{$\chi^2$/$dof$} & \colhead{$\gamma$} & \colhead{log($f_{\rm b}$(Hz))}  & \colhead{$\beta$} & \colhead{$A$}  }
\startdata
\multicolumn{6}{c}{Sharply-Broken Power-Law Model Fits} \\ \hline
T & 207.5/178 & 1.48$\pm$0.14          & --3.36$\pm$0.04 & 3.00$\pm$0.12          & 0.95$\pm$0.03 \\ 
S & 217.9/178 & 1.56$\pm$0.18          & --3.55$\pm$0.09 & 2.56$^{+0.16}_{-0.14}$ & 1.08$\pm$0.04 \\ 
M & 173.0/178 & 1.54$\pm$0.14          & --3.34$\pm$0.05 & 3.10$\pm$0.18          & 0.91$\pm$0.03 \\ 
H & 178.0/178 & 1.28$\pm$0.12          & --3.29$\pm$0.04 & 3.02$\pm$0.16          & 1.01$\pm$0.03  \\ \hline 
T & 208.6/180 & 1.46 (fixed)           & --3.39$\pm$0.05 & 2.89 (fixed)           & 0.99$\pm$0.03 \\ 
S & 222.6/180 & 1.46 (fixed)           & --3.48$\pm$0.06 & 2.89 (fixed)           & 1.00$\pm$0.04 \\ 
M & 175.1/180 & 1.46 (fixed)           & --3.40$\pm$0.05 & 2.89 (fixed)           & 1.03$\pm$0.03 \\ 
H & 187.2/180 & 1.46 (fixed)           & --3.27$\pm$0.05 & 2.89 (fixed)           & 0.92$\pm$0.03 \\ \hline
T & 207.6/180 & 1.46 (fixed)           & --3.36 (fixed)& 3.00$\pm$0.14          & 0.94$\pm$0.03 \\
S & 231.6/180 & 1.46 (fixed)           & --3.36 (fixed)& 2.94$\pm$0.24          & 0.73$\pm$0.04  \\
M & 173.4/180 & 1.46 (fixed)           & --3.36 (fixed)& 3.08$^{+0.22}_{-0.16}$ & 0.97$\pm$0.03 \\
H & 192.5/180 & 1.46 (fixed)           & --3.36 (fixed)& 2.76$\pm$0.14          & 1.10$\pm$0.03 \\ \hline
\multicolumn{6}{c}{Slowly-Broken Power-Law Model Fits} \\ \hline
T & 210.8/178 & 1.26$\pm$0.12          & --3.34$\pm$0.05 & 3.34$\pm$0.12          & --3.10$\pm$0.03 \\ 
S & 218.3/178 & 1.04$\pm$0.24          & --3.77$\pm$0.09 & 2.68$\pm$0.12          & --2.16$\pm$0.04 \\ 
M & 172.9/178 & 1.42$\pm$0.12          & --3.23$\pm$0.06 & 3.72$\pm$0.22          & --3.73$\pm$0.03 \\ 
H & 186.4/178 & 1.16$\pm$0.10          & --3.20$\pm$0.04 & 3.58$\pm$0.20          & --2.70$\pm$0.03 \\ \hline
T & 210.9/180 & 1.21 (fixed)           & --3.36$\pm$0.05 & 3.33 (fixed)           & --2.91$\pm$0.04 \\ 
S & 225.5/180 & 1.21 (fixed)           & --3.44$\pm$0.07 & 3.33 (fixed)           & --3.03$\pm$0.03 \\ 
M & 174.5/180 & 1.21 (fixed)           & --3.39$\pm$0.05 & 3.33 (fixed)           & --2.86$\pm$0.03 \\ 
H & 188.3/180 & 1.21 (fixed)           & --3.23$\pm$0.04 & 3.33 (fixed)           & --2.85$\pm$0.03 \\ \hline
T & 211.0/180 & 1.21 (fixed)           & --3.19 (fixed) & 3.36$\pm$0.14          & --2.92$\pm$0.03 \\ 
S & 229.6/180 & 1.21 (fixed)           & --3.19 (fixed) & 3.30$\pm$0.26          & --3.13$\pm$0.04 \\
M & 174.3/180 & 1.21 (fixed)           & --3.19 (fixed) & 3.46$\pm$0.20          & --2.90$\pm$0.03 \\
H & 195.3/180 & 1.21 (fixed)           & --3.19 (fixed) & 3.08$\pm$0.16          & --2.76$\pm$0.03 \\
\enddata
\tablecomments{$A$ is the PSD amplitude in units of Hz$^{-1}$ at the break frequency $f_{\rm b}$.}
\end{deluxetable}

\begin{deluxetable}{lccccccc}
\tabletypesize{\footnotesize}
\tablewidth{6.5in}
\tablenum{3}
\tablecaption{PSD Model Fits Using a Sharply-Broken Power-law plus a Lorentzian \label{tab3}}
\tablehead{
\colhead{Band} & \colhead{$\chi^2$/$dof$} & \colhead{$\gamma$} & \colhead{log($f_{\rm b}$(Hz))}  & \colhead{$\beta$} & \colhead{$A$}  & \colhead{log($f_{\rm L}$)} &  \colhead{$R$ ($\%$)}  }
\startdata   
T & 212.7/178 &  1.46 (fixed) &  --3.60$^{+0}_{-0.06}$    & 2.68$\pm$0.10 &  1.33$\pm$0.03 &  --3.34$\pm$0.04   &   3.0$\pm$0.5 \\         
S & 218.2/178 &  1.46 (fixed) &  --3.60$^{+0}_{-0.08}$    & 2.54$\pm$0.14 &  1.19$\pm$0.04 &  --3.34$\pm$0.04   &  $<$2.0   \\
M & 178.1/178 &  1.46 (fixed) &  --3.60$^{+0}_{-0.06}$    & 2.70$\pm$0.14 &  1.35$\pm$0.03 &  --3.34$\pm$0.04   &  3.5$^{+0.5}_{-1.0}$ \\  
H & 193.7/178 &  1.46 (fixed) &  --3.60$^{+0}_{-0.06}$    & 2.42$\pm$0.12 &  1.38$\pm$0.03 &  --3.34$\pm$0.04   &  5.0$\pm$0.5   \\
\enddata
\tablecomments{Results to fitting the PSDs with a model consisting of a sharply-broken power-law
plus a Lorentzian. $A$ is the PSD amplitude in units of Hz$^{-1}$ at the break frequency $f_{\rm b}$.
The break frequency $f_{\rm b}$ was constrained to be 10$^{-3.60}$ Hz or lower.
The low-frequency slope $\gamma$ was fixed at --1.46.
Based on the sum of the $\chi^2$ values for the S, M and H bands, the best-fit
value of the Lorentzian centroid frequency $f_{\rm L}$ was found to be 10$^{-3.34 \pm 0.04}$ Hz;
fit parameters above correspond to that value of $f_{\rm L}$.
In all fits, the quality factor of the Lorentzian $Q$ was assumed to be 5.}
\end{deluxetable}

%%%%%%%%%%%%%%%%%%%%%%%%%%%%%%%%%%%%%%%%%%%%%%%% FIGURES %%%%%%%%%%%%%%%
%%%%\clearpage

\begin{figure}        %%%%%% figure 1 = L.C.'s. Plot 2005 data only.
\epsscale{0.95}
\plotone{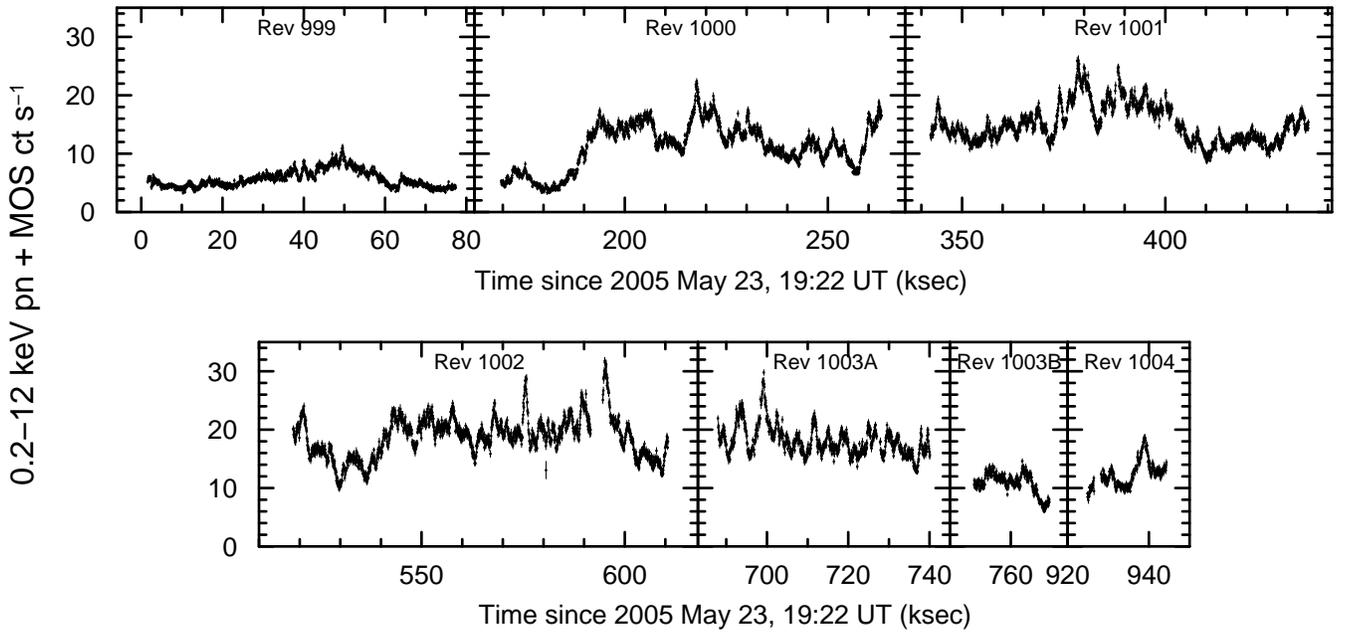}
\caption{0.2--12 keV EPIC light curves, summed over the 
pn and MOS cameras, binned to 60 s, 
for the 2005 observation. All panels have the same scale.}
\end{figure}

\begin{figure}        %%%%%% figure 2 = rms spec
\epsscale{0.95}
\plotone{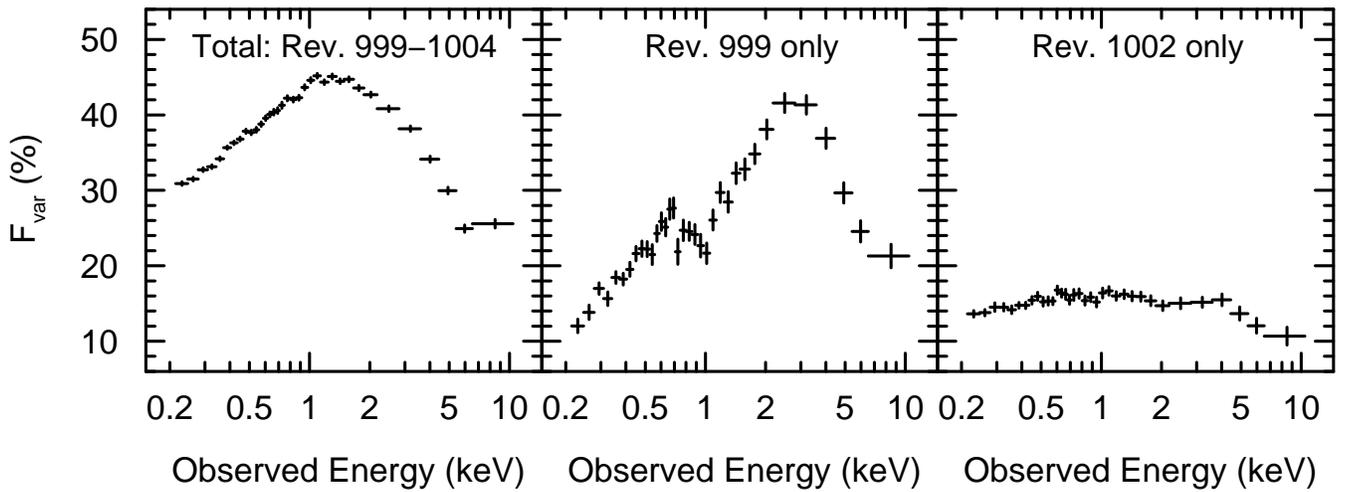}
\caption{$\fv$ spectra, constructed using EPIC-pn data,
for the entire 2005 observation and for the lowest- and highest-flux
revolutions.}
\end{figure}

\begin{figure}    %%%%%% figure 3 = rms-flux relation
\epsscale{0.50}
\plotone{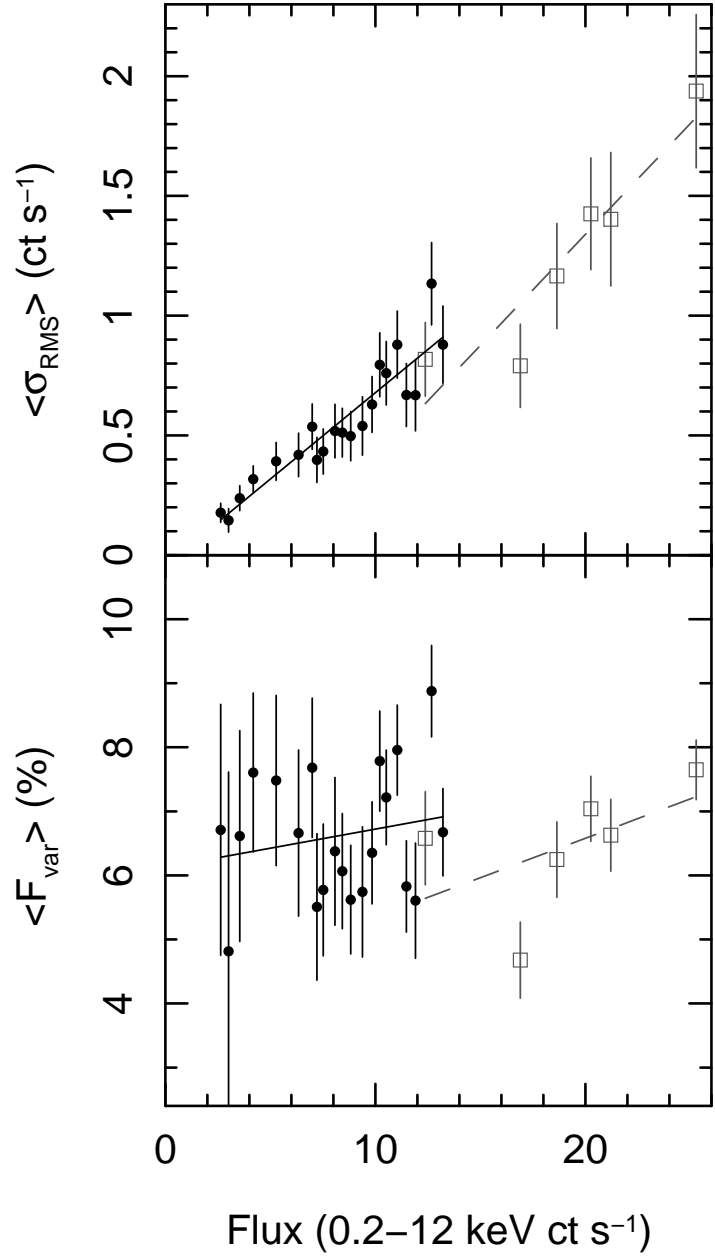}
\caption{The 2005 pn data (black circles) and the 2001 pn
data (gray open squares) are both consistent with
a linear relation between rms variability amplitude (top)
and with constant average fractional variability amplitude
$<$$\fv$$>$. The black solid and gray dashed lines are the
best-fit relations for the 2005 and 2001 data, respectively.}
\end{figure}

\begin{figure}         %%%%% figure 4 = T-band PSD
\epsscale{0.70}
\plotone{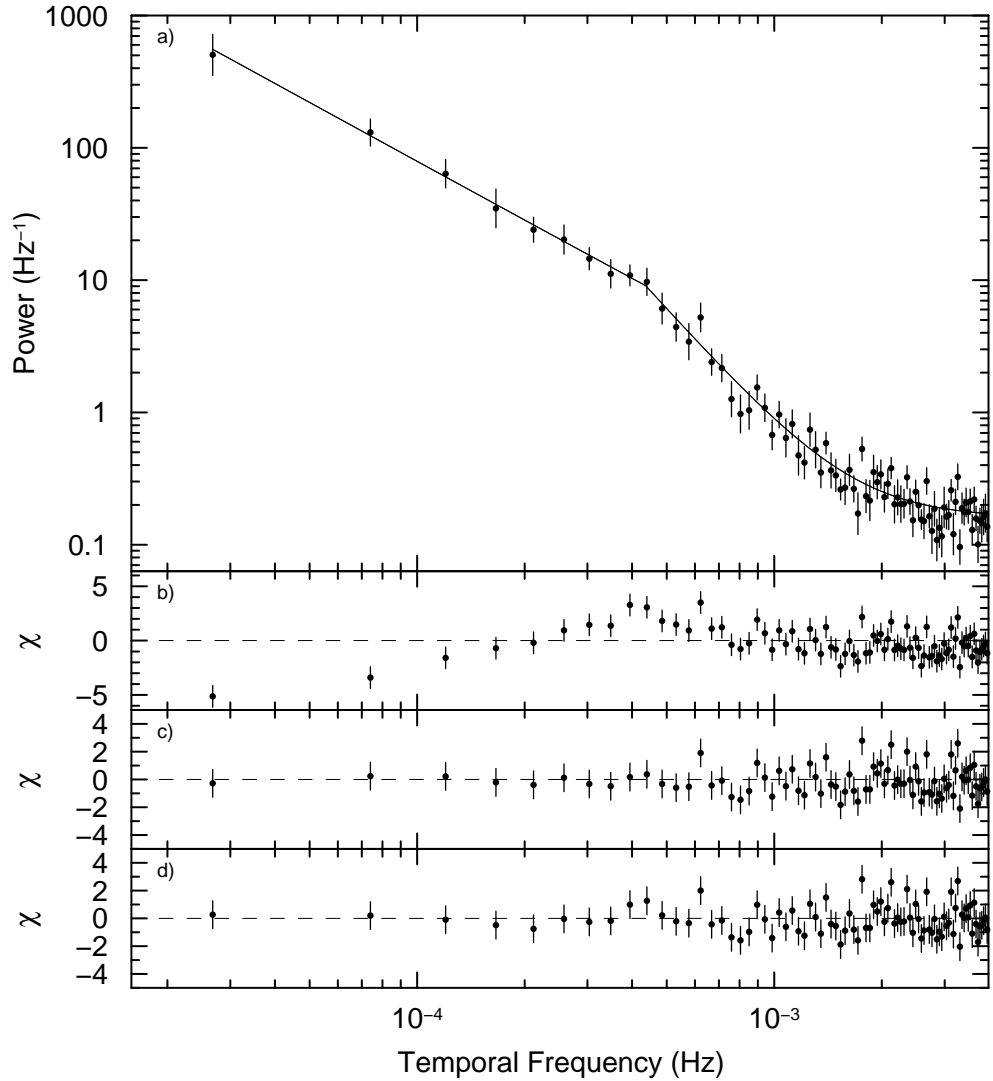}
\caption{{\it a)} Power density spectrum for the T-band light curve.
The solid line denotes the best-fit sharply-broken power-law fit.
{\it b)} Residuals to the best-fit unbroken power-law model;
residuals are measured in log space, e.g., based on the logs of the 
PSD model, data and errors.
{\it c)} Residuals to the best-fit sharply-broken power-law model.
{\it d)} Residuals to the best-fit slowly-bending power-law model. Power due
to Poisson noise was included in the models in all panels.}
\end{figure}

\begin{figure}   %%%% figure 5 = SMH band PSD for SIN and SLO
\epsscale{0.70}
\plotone{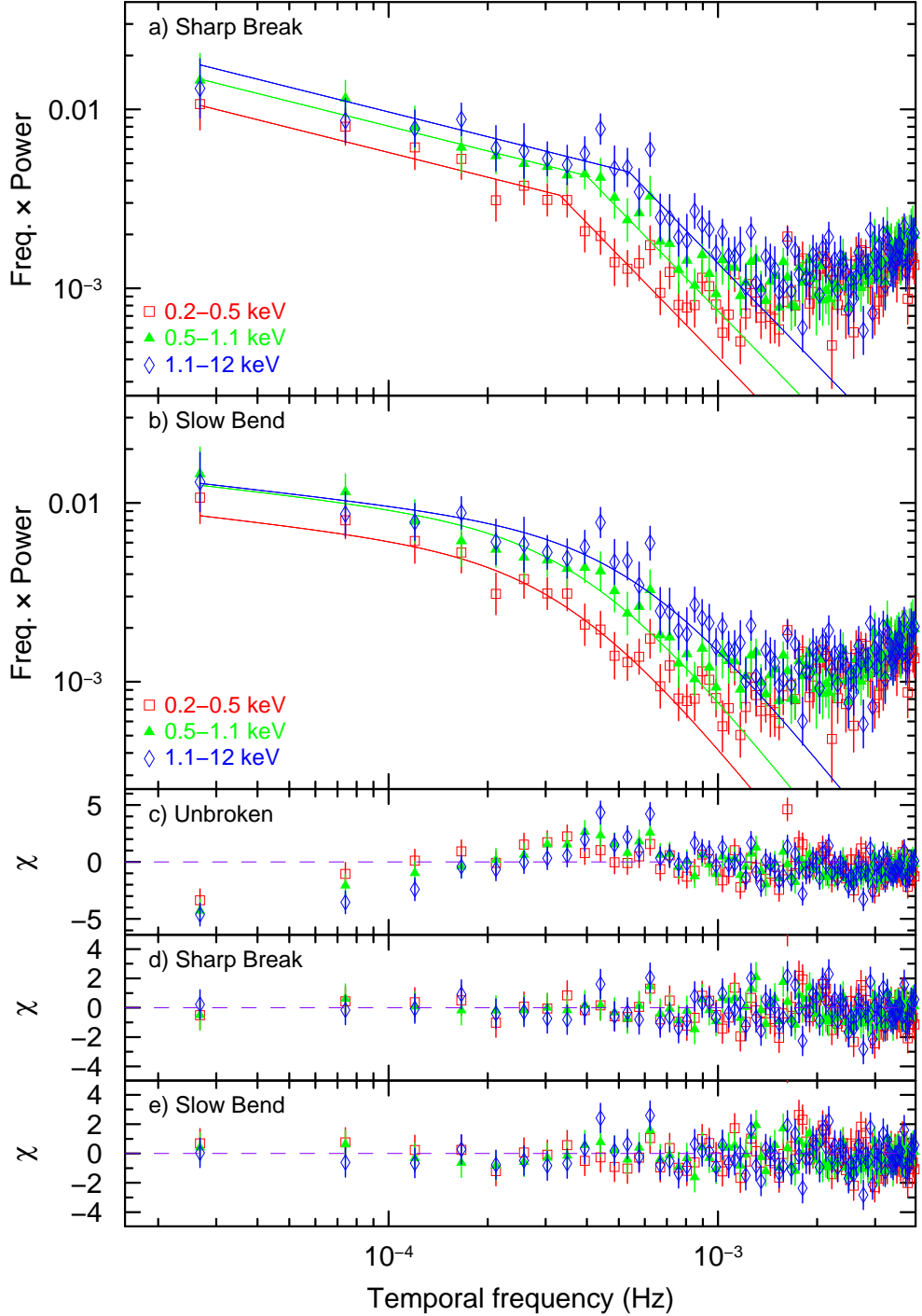}
\caption{{\it a)} Power density spectra for the S, M, and H-band light curves,
plotted in $f - f\times$$P(f)$ space.
The solid lines show the best-fit sharply-broken power-law model assuming
the power-law slopes above and below the break $\beta$ and $\gamma$ are equal in all bands.
The models are plotted with the constant level of power due to Poisson noise omitted for clarity. 
{\it b)} Same as {\it a)}, but the solid lines show the best-fit slowly-bending power-law model assuming
the power-law slopes above and below the break $\beta$ and $\gamma$ are equal in all bands.
{\it c)} Residuals to the best-fit unbroken power-law model.
{\it d)} Residuals to the best-fit sharply-broken power-law model.
{\it e)} Residuals to the best-fit slowly-bending power-law model (power due
to Poisson noise was included in the models in panels {\it b)}--{\it d)}.}
\end{figure}

\begin{figure}     %%%%% figure 6 = QPO plot
\epsscale{0.70}
\plotone{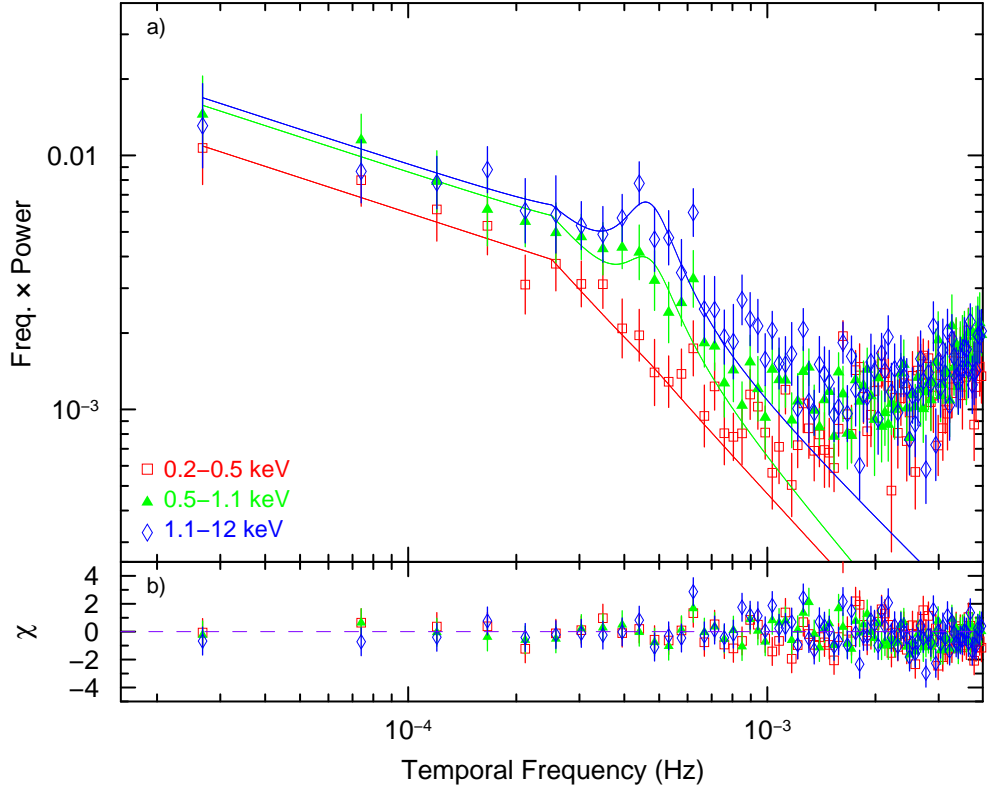}
\caption{{\it a)} Power density spectra for the S, M, and H-band light curves,
plotted in $f - f\times$$P(f)$ space.
The solid lines show the best-fit model consisting of
a sharply-broken power-law, with break frequency constrained 
to be 10$^{-3.60}$ Hz or lower, plus a Lorentzian
with a centroid frequency of 10$^{-3.32}$ Hz and quality factor $Q = 5$.
The models are plotted with the constant level of power due to Poisson noise omitted for clarity. 
{\it b)} Residuals to the best-fit model (with power due to
Poisson noise included).}
\end{figure}

\begin{figure}    %%%%% figure 7 = QPOCONTOUR  R versus F_L
\epsscale{0.90}
\plotone{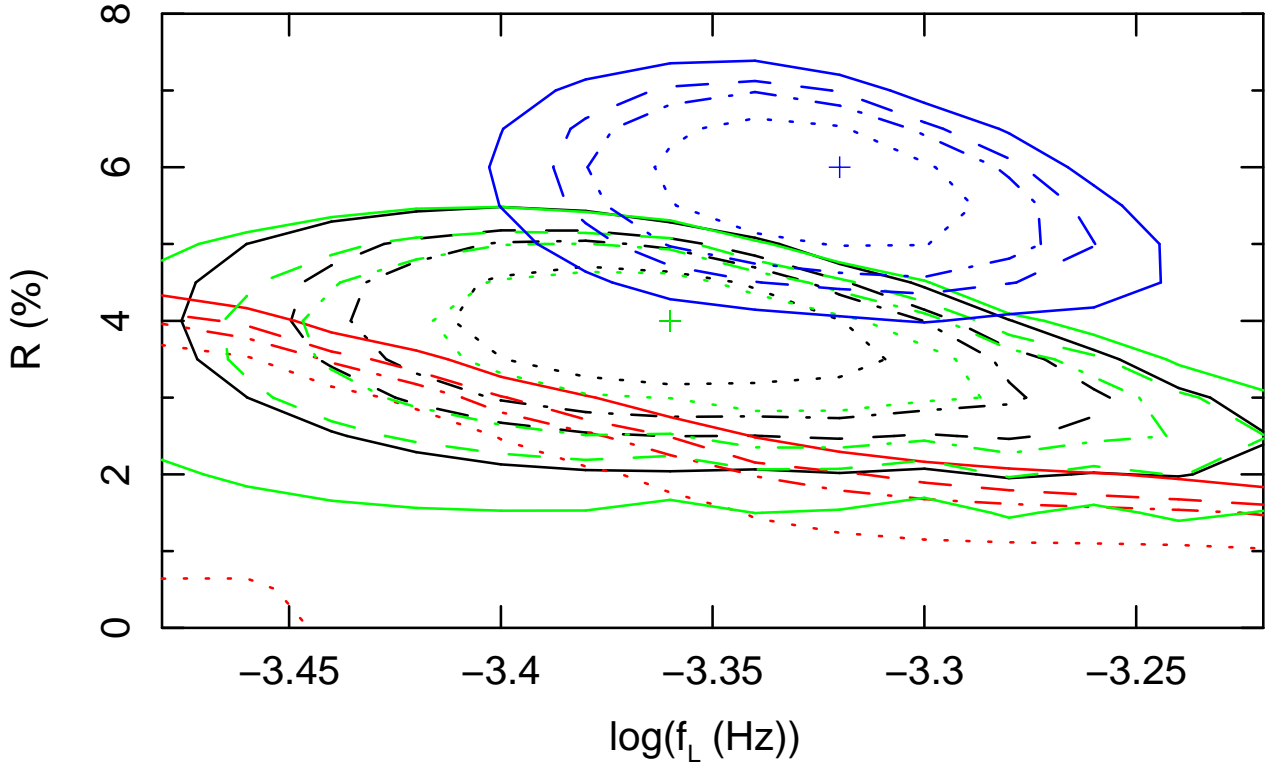}
\caption{Derived confidence contours of the Lorentzian amplitude $R$ versus centroid frequency $f_{\rm L}$
for each band; black, red, green, and blue denote T, S, M and H bands, respectively.
68, 90, 95 and 99$\%$ contour levels are shown (solid, dashed, dash-dotted, and dotted, respectively).
The blue cross marks the best-fit value for the H band.
The green cross marks the best-fit value for both the M and T bands.}
\end{figure}

\begin{figure}    %%%%% figure 8 = COH 
\epsscale{0.90}
\plotone{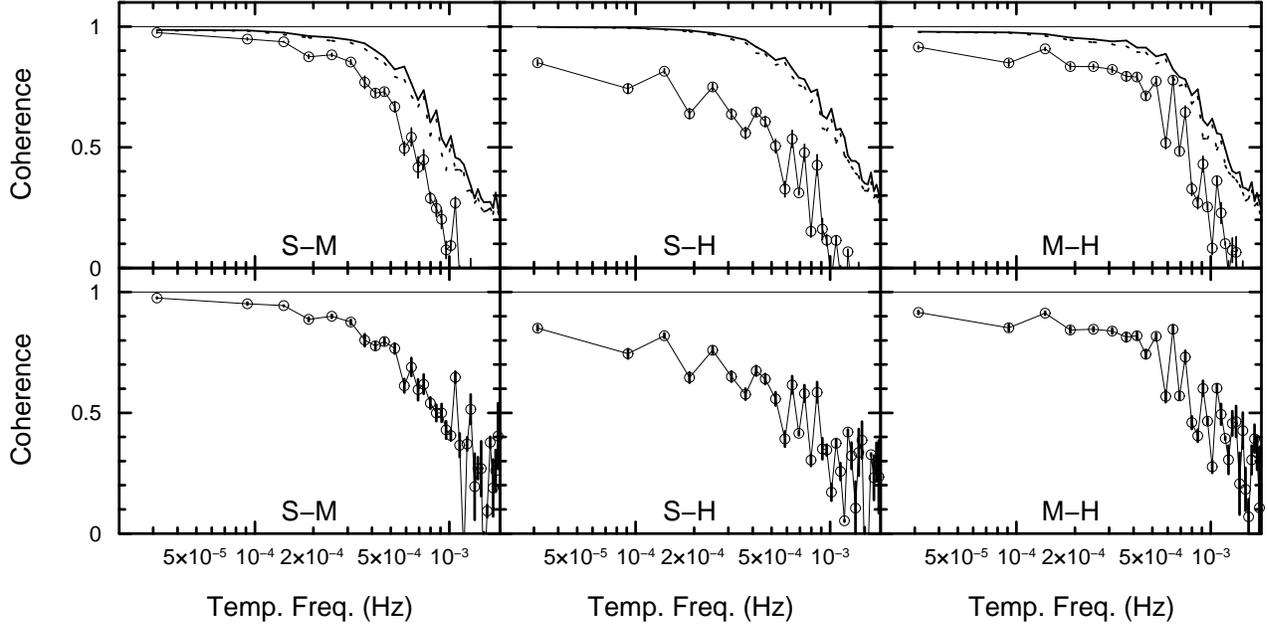}
\caption{The top row shows the observed coherence functions for S--M, S--H and M--H
light curve pairs. The thin solid line denotes unity coherence.
The dashed and dotted lines denote, respectively, the 90$\%$ and 
95$\%$ confidence limits for spurious lack of coherence due
to the combination of Poisson noise and the effect 
of differing PSD shapes as determined by Monte Carlo simulations.
The bottom row shows the estimate of the ``intrinsic'' coherence (see text for details).}
\end{figure}

\begin{figure}    %%%%%% fig 9 = PHI
\epsscale{0.90}
\plotone{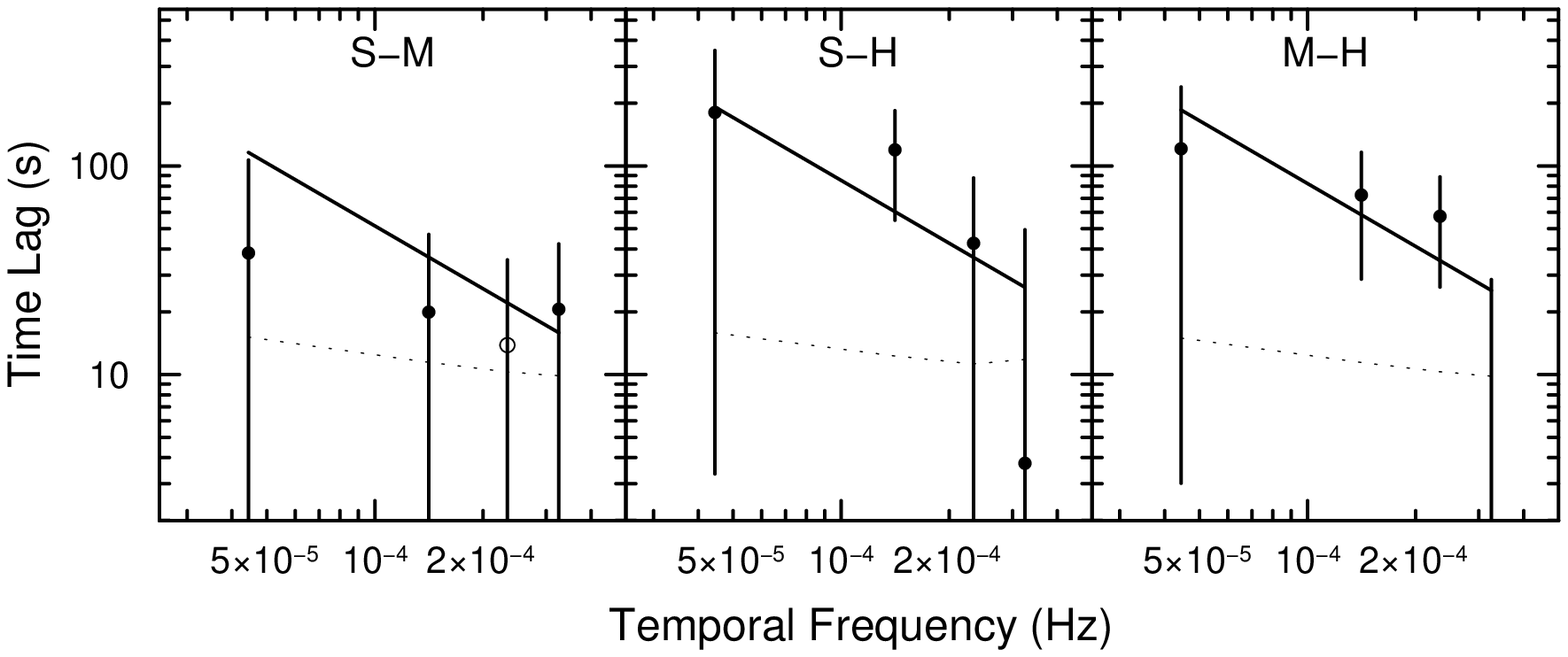}
\caption{Time lag spectra for Mkn 766.
Positive lags denote soft band variations leading those in
the hard band. A negative lag was detected in the
second-highest temporal frequency bin in the S--M time lag spectrum, 
and its magnitude is denoted by an open circle.
The dashed lines denotes the sensitivity limits
for time lag detection; see text for details.
The solid lines denote the best-fit relations
to $\tau(f) = Cf^{-1}$.}
\end{figure}

%%%%%%%%%%%%%%%%%%%%%%%%%%%%%%%%%%%%%% APPENDIX %%%%%%%%%%%%%%%%%%%%
\appendix

\section{PSD consistency}

We tested for strong non-stationarity in the form of a change in 
the shape of the underlying PSD, between the 2001 and 2005 PSDs 
and between revolutions in the 2005 long-look. For two PSDs 
observing the same stationary process at two different times,
the only difference in PSD values at a given temporal frequency 
should be consistent with the expected scatter.  The scatter in the 
measured PSD about the intrinsic, underlying PSD follows a $\chi^2$ 
distribution with two degrees of freedom (e.g., Priestley 1981);  
Papadakis \& Lawrence (1995) thus outlined a method to test for 
non-stationarity, defining a statistic $S$ to quantify
differences between two PSDs. For a stationary process, $S$ will
be distributed with a mean of 0 and a variance equal to 1 (see also VFN03).

We considered PSDs for each revolution from 999 to 1003A 
using the summed pn + MOS1+MOS2 T, S, M and H-band light 
curves as well as for the summed pn + MOS2 light curve from 2001.  
Light curves were binned to 60 seconds and truncated to a common 
duration of 52.6 ksec; PSDs were generated by binning periodogram 
points by a factor of 25.

The $S$ statistic was calculated by summing over temporal frequencies 
$<$10$^{-2.6}$ Hz to exclude PSD points that 
were dominated by power due to Poisson noise. For all PSD pairs and 
bands, $\vert$$S$$\vert$ was found to 0.54 or less. Extending 
the duration to 74.0 ksec to compare revolutions 265 and 999--1002
and extending the duration to 92.2 ksec to compare revolutions
265 and 1000--2 also yielded similarly small values of $\vert$$S$$\vert$ 
($\lesssim$0.5) for all bands.
To summarize, all values of $\vert S \vert$ were less than 1
for all light curve pairs and bands;
we therefore have found no evidence at 1$\sigma$ confidence or greater
for strong non-stationarity in the
PSDs of Mkn 766 at any band from one revolution to the next in 
the 2005 long-look, or between 2001 and 2005.

\end{document}